\documentclass[review,12pt]{elsarticle}

\usepackage{amssymb}
\usepackage{amsthm}
\usepackage{amsmath}
\usepackage{mathrsfs}
\usepackage{graphicx}
\usepackage{array}
\usepackage{epstopdf}
\usepackage{float}
\usepackage{caption}
\usepackage{subcaption}
\usepackage{bm}
\usepackage{hyperref}
\usepackage{bbm}
\usepackage{mathrsfs}
\usepackage{cleveref}
\usepackage{booktabs}
\usepackage{color, colortbl}
\definecolor{Gray}{gray}{0.9}
\usepackage{soul}
\usepackage{color,soul} 
\usepackage{color} 
\usepackage{siunitx}
\usepackage[margin=2.5cm]{geometry}
\biboptions{sort&compress}
\graphicspath{ {./images/} }

\newcolumntype{F}[1]{%
    >{\raggedright\arraybackslash\hspace{0pt}}p{#1}}%
\newcolumntype{T}[1]{%
    >{\centering\arraybackslash\hspace{0pt}}p{#1}}%
    
\journal{Journal of the Mechanics and Physics of Solids}

\date{\today}

\makeatletter
\def\@author#1{\g@addto@macro\elsauthors{\normalsize%
    \def\baselinestretch{1}%
    \upshape\authorsep#1\unskip\textsuperscript{%
      \ifx\@fnmark\@empty\else\unskip\sep\@fnmark\let\sep=,\fi
      \ifx\@corref\@empty\else\unskip\sep\@corref\let\sep=,\fi
      }%
    \def\authorsep{\unskip,\space}%
    \global\let\@fnmark\@empty
    \global\let\@corref\@empty  
    \global\let\sep\@empty}%
    \@eadauthor={#1}
}
\makeatother

\begin{document}

\begin{frontmatter}

\title{Inverse design and additive manufacturing of shape-morphing structures based on functionally graded composites}

\author{Hirak Kansara \fnref{QMUL}}
\author{Mingchao Liu\corref{cor1}\fnref{NTU}}\ead{mingchao.liu@ntu.edu.sg}
\author[Nottingham,Nottingham2]{Yinfeng He} 
\author{Wei Tan\corref{cor1}\fnref{QMUL}}\ead{wei.tan@qmul.ac.uk}

\address[QMUL]{School of Engineering and Materials Science, Queen Mary University of London, London, E1 4NS, UK}

\address[NTU]{School of Mechanical and Aerospace Engineering, Nanyang Technological University, Singapore, 639798, Republic of Singapore}

\address[Nottingham]{Nottingham Ningbo China Beacons of Excellence Research and Innovation Institute, University of Nottingham Ningbo China, Ningbo, China}

\address[Nottingham2]{Centre for Additive Manufacturing, Faculty of Engineering, University of Nottingham, Nottingham NG8 1BB, UK}

\cortext[cor1]{Corresponding authors.}

\begin{abstract}

Shape-morphing structures possess the ability to change their shapes from one state to another, and therefore, offer great potential for a broad range of applications. A typical paradigm of morphing is transforming from an initial two-dimensional (2D) flat configuration into a three-dimensional (3D) target structure. One popular fabrication method for these structures involves programming cuts in specific locations of a thin sheet material (i.e.~kirigami), forming a desired 3D shape upon application of external mechanical load. By adopting the non-linear beam equation, an inverse design strategy has been proposed to determine the 2D cutting patterns required to achieve an axisymmetric 3D target shape. Specifically, tailoring the localised variation of bending stiffness is the key requirement. In this paper, a novel inverse design strategy is proposed by modifying the bending stiffness via introducing distributed modulus in functionally graded composites (FGCs). To fabricate the FGC-based shape-morphing structures, we use a multi-material 3D printer to print graded composites with voxel-like building blocks. The longitudinal modulus of each cross-sectional slice can be controlled through the rule of mixtures according to the micro-mechanics model, hence matching the required modulus distribution along the elastic strip. Following the proposed framework, a diverse range of structures is obtained with different Gaussian curvatures in both numerical simulations and experiments. A very good agreement is achieved between the measured shapes of morphed structures and the targets. In addition, the compressive rigidity and specific energy absorption during compression of FGC-based hemi-ellipsoidal morphing structures with various aspect ratios were also examined numerically and validated against experiments. By conducting systematical numerical simulations, we also demonstrate the multifunctionality of the modulus-graded shape-morphing composites. For example, they are capable of blending the distinct advantages of two different materials, i.e. one with high thermal (but low electrical) conductivity, and the other is the other way around, to achieve combined effective properties in a single structure made by FGCs. This new inverse design framework provides an opportunity to create shape-morphing structures by utilising modulus-graded composite materials, which can be employed in a variety of applications involving multi-physical environments. Furthermore, this framework underscores the versatility of the approach, enabling precise control over material properties at a local level.
\end{abstract}


\end{frontmatter}

\section{Introduction}
\label{section:Introduction}
Shape-morphing structures that can change their shapes from one form to another are offering great potential compared to conventional engineering structures \cite{oliver2016morphing,liu2023snap}. These types of structures have found widespread applications ranging from soft robotics \cite{ kotikian2019untethered,liu2021robotic}, biomedical devices \cite{zhou20204d}, deployable space structures \cite{mccue2021controlled}, mechanical metamaterials \cite{neville2016shape, dudek2022micro}, aircraft aerodynamic control systems \cite{li2018review} and infrastructure constructions \cite{li2021shape}. 

Developing shape-morphing structures that can transform from two-dimensional (2D) flat sheets into three-dimensional (3D) structures is desired due to the simplicity of fabrication and transportation. However, achieving structures with non-zero Gaussian curvature ($\kappa \neq 0$) using thin flat sheets (with $\kappa = 0$) is difficult. According to the \emph{Gauss' Theorema Egregium} \cite{gauss1828disquisitiones}, a deformation that does not change the length or area (isometries) cannot change the Gaussian curvature of the surface. This limits the possible target structures that could be constructed using flat plates. To overcome such challenge, the deformation of flat surfaces must be coupled with localised stretching/shrinkage. This can be achieved using responsive materials (e.g. hydrogel \cite{kirillova20174d}, shape memory polymers \cite{behl2009shape}, elastomers \cite{tang2022stretch}, etc.) subjected to external stimuli, including pneumatic \cite{siefert2019bio}, photonic \cite{kim2019light}, thermal \cite{aharoni2018universal}, chemical \cite{nojoomi2018bioinspired}, as well as electric \cite{hajiesmaili2019reconfigurable, chortos20203d} and magnetic \cite{bastola2021shape,wang2021evolutionary, yarali2022magneto} fields. These materials with localised actuation-induced expansion/contraction enable changes in the Gaussian curvature of surfaces. However, such approaches usually employ complex fabrication methods \cite{naficy20174d,wang2018programmable}, diminishing the mass-production scalability of shape-morphing structures. Additionally, changes in the length of materials can induce significant mechanical strain, which may be undesired in some applications (e.g. flexible electronic device \cite{zhang2017printing} and deployable solar panels \cite{chen2019autonomous}).

An alternate approach is to remove excess material by introducing distributed cuts prior to assembly, i.e.~kirigami \cite{neville2016shape, choi2019programming, jin2020kirigami}. By programming cuts in specific locations, structures with different Gaussian curvatures can be generated. This curvature should be referred to as the `Apparent Gaussian Curvature' (AGC), whereby the Gaussian curvature locally remains zero even though on a global scale it has changed for the 3D structure \cite{liu2020tapered}. This strategy of making cuts divides a complete sheet into several facets connected to a central hub, and therefore, prevents the material from experiencing undesired excess strain on each facet. Consequently, making it an ideal method to produce flexible devices where little strain is tolerable \cite{zhang2017printing,chen2019autonomous}. Nevertheless, the problem of programming cuts on planar sheets that form desired 3D structures remains. Traditionally, the approach to such inverse design problems has been accommodated by an arduous trial-and-error process. Numerical optimisation techniques have also been successfully employed \cite{xu2019optimization,qin2020genetic}. However, analytical approaches to address this type of inverse design problem have only been developed in the recent past \cite{liu2020tapered,fan2020inverse}.

The inverse design strategy established by Liu et al.~\cite{liu2020tapered}, for determining 2D cut patterns to form desired 3D axisymmetric structures, is based on the theory of tapered elastica \cite{stuart2000buckling}. Elastica is defined as a thin strip (beam) made from elastic materials, while a tapered one indicates a strip with non-uniform geometry (i.e.~width and thickness). The axisymmetric structure is combined using multiple strips that are radially connected to a central hub, in which the deformation of each strip under mechanical loading can be determined by the elastica theory. By extension, an equation relating the shape of the strip and the curvature of a 3D shape formed by applying load can therefore be derived. According to Liu et al.~\cite{liu2020tapered}, for strips with uniform thickness, the local bending stiffness can be manipulated through changes in width. This results in the occurrence of large gaps between adjacent strips in the morphed structure. To create tessellated morphing structures (i.e.~no gaps between adjacent strips in the morphed state), the geometric constraint of tessellation needs to be satisfied. One possible way is to taper both the width and thickness of the 2D flat sheets \cite{liu2020tapered}. Realising the morphing mechanism, changing the width and thickness, and by extension, the moment of inertia, the local bending stiffness can be manipulated \cite{lee1993elastica}. With advances in 3D printing, structures with varying thicknesses can be printed but may not be compatible with brittle materials. An approach proposed in a recent study by Zhang et al.~\cite{zhang2022shape} on morphing structures investigated the effect of introducing distributed local porosity on the deformation of tapered strips, which can be fabricated through laser cutting of an elastic sheet, but may diminish the load-bearing capacity of the structure due to removal of material. Consequently, alternative methods of attaining the required local bending stiffness must be explored. 

Nevertheless, previous studies to reach the required bending stiffness relied on changes in the geometry of the strips and assumed uniform Young's modulus. Upon inspecting the formula of the bending stiffness, i.e.~$B = f(I, E)$, it can be realised that $B$ is also dependent upon the Young's modulus ($E$), not just the moment of inertia ($I$). Through variation in the Young's modulus along position, desired distribution of the bending stiffness can also be achieved. The exact modulus distribution required can readily be implemented into Finite Element Method (FEM) simulations, but, the challenge lies in the fabrication of such functionally graded materials. {Cheng et al. \cite{cheng2023programming} recently used photolithography to manufacture graded microlattice structures for shape-morphing structures. The Young's modulus of stretch-dominated triangular microlattice is controlled by its local relative density, similar to the perforated structure proposed by Zhang et al.~\cite{zhang2022shape}.} The development of additive manufacturing technologies, such as 3D printing \cite{kennedy2020printing}, provides another potential solution, but such technology with limited control of the Young's modulus of \emph{in situ} mixed materials faces challenges to match the exact graded modulus profile.  Nevertheless, inspired by recent works on Functionally Graded Composites (FGCs) using multi-material volumetric pixel-based (i.e.~voxel-based) 3D printing \cite{ituarte2019design, mirzaali2020mechanics, he2021exploiting, skylar2019voxelated}, a novel paradigm for the inverse design and fabrication of modulus-graded shape-morphing structures based on FGCs is presented in this work. 

By discretising the graded strips with geometric and physical parameters obtained from a theoretical model using voxels, soft and rigid phases can be allocated to each voxel, facilitating the development of tailored FGCs with precisely controlled mechanical properties \cite{loh2018overview}. By considering cross-sectional slices, along the length of the strip, the longitudinal modulus of each slice can be controlled through the rule of mixtures, with voxels being assigned materials based on the required volume fraction. One requirement to manufacture FGCs is to have a good interface cohesion between soft and rigid materials, to avoid failure mechanisms prevalent in traditional composites such as debonding, delamination, and matrix cracking \cite{talreja2023damage}. In addition, applications such as those in the aerospace and automobile industries may require a structure with high stiffness and strength to withstand external loading \cite{xu2018review}. Consequently, the load-bearing capacities of shape-morphing FGCs are assessed experimentally and numerically through indentation (i.e. by applying compressive loads).

Furthermore, the proposed approach of creating shape-morphing structures using modulus-graded composites being composed of two or more phases, each with unique properties, which can be combined to achieve superior effective mechanical and physical properties of the overall structure. This results in multifunctional FGCs  that allow designers and engineers to create structures with multiple desired functionalities such as shape-morphing, energy absorption, and vibration damping in a single structure. Additionally, through the choice of materials, such composites can be designed to have specific combinations of properties, such as thermal and electrical conductivity, which facilitates an application-specific approach when designing FGCs-based shape-morphing structures with multifunctionalities, such as electromagnetic shielding, energy storage, and sensing capabilities.

In this work, an inverse design framework based on graded elastica theory and composite micromechanics together with an additive manufacturing strategy is presented to design axisymmetric shape-morphing composite structures. The main novel contributions herein are: (i) For the first time, a 3D tessellated shape-morphing structure based on FGCs with the graded-modulus profile is proposed and validated by both FEM simulations and physical experiments. (ii) A voxel-based multi-materials additive manufacturing method is developed to fabricate the FGC-based shape-morphing structures. (iii) We also explore the load-bearing capacities (both the rigidity and the specific energy absorption) of the 3D morphed composite structures (iv) We further demonstrate the multifunctionality of shape-morphing FGCs by simulating the heat transfer and the electric potential induced by a passing current, aided by suitable material selection. This approach enables us to leverage the optimal properties of each material. Overall, our framework opens new opportunities for the efficient and cost-effective design and manufacturing of shape-morphing structures using composite materials.

\section{Mechanics of functionally graded composite beams}
\label{section:mechanics of FGC beam}

\subsection{Graded elastica model}
\label{section:theory rectangular strip}

Following the theoretical framework developed by Liu et al.~\cite{liu2020tapered}, inverse design problems for axisymmetrical shape-morphing structures can be addressed using the tapered elastica theory. The underlying principle of morphing to a desired 3D shape relies on the manipulation of mechanical properties, i.e.~the bending stiffness, of the 2D flat sheet. This is done by either changing the moment of inertia, $I(s)$, i.e.~width/thickness, or the local Young's modulus, $E(s)$. However, it may not be feasible to vary the thickness and only varying the width would result in large gaps \cite{liu2020tapered}. Alternatively, by introducing non-uniform Young's modulus that varies along the length of the geometry, similar alteration of the bending stiffness $B(s)$ can be achieved,

\begin{equation}
    B(s) = E(s) I(s),
\end{equation}
where $I(s) = w(s) t^3_0/12$ is the non-uniform moment of inertia. Note that here we only keep width $w(s)$ as a variable parameter to satisfy the tessellation condition (which will be discussed later), whilst constraining the thickness to a constant value $t_0$.

By accounting for the non-uniform distribution of Young's modulus, the tapered elastica equation can be modified, which is therefore referred to as graded elastica. Nonetheless, the focus of this study is to solely obtain the modulus distribution which is the only unknown, otherwise, there would exist multiple solutions for different configurations.

\begin{figure}[ht]
   \centering
   \includegraphics[width=1\textwidth]{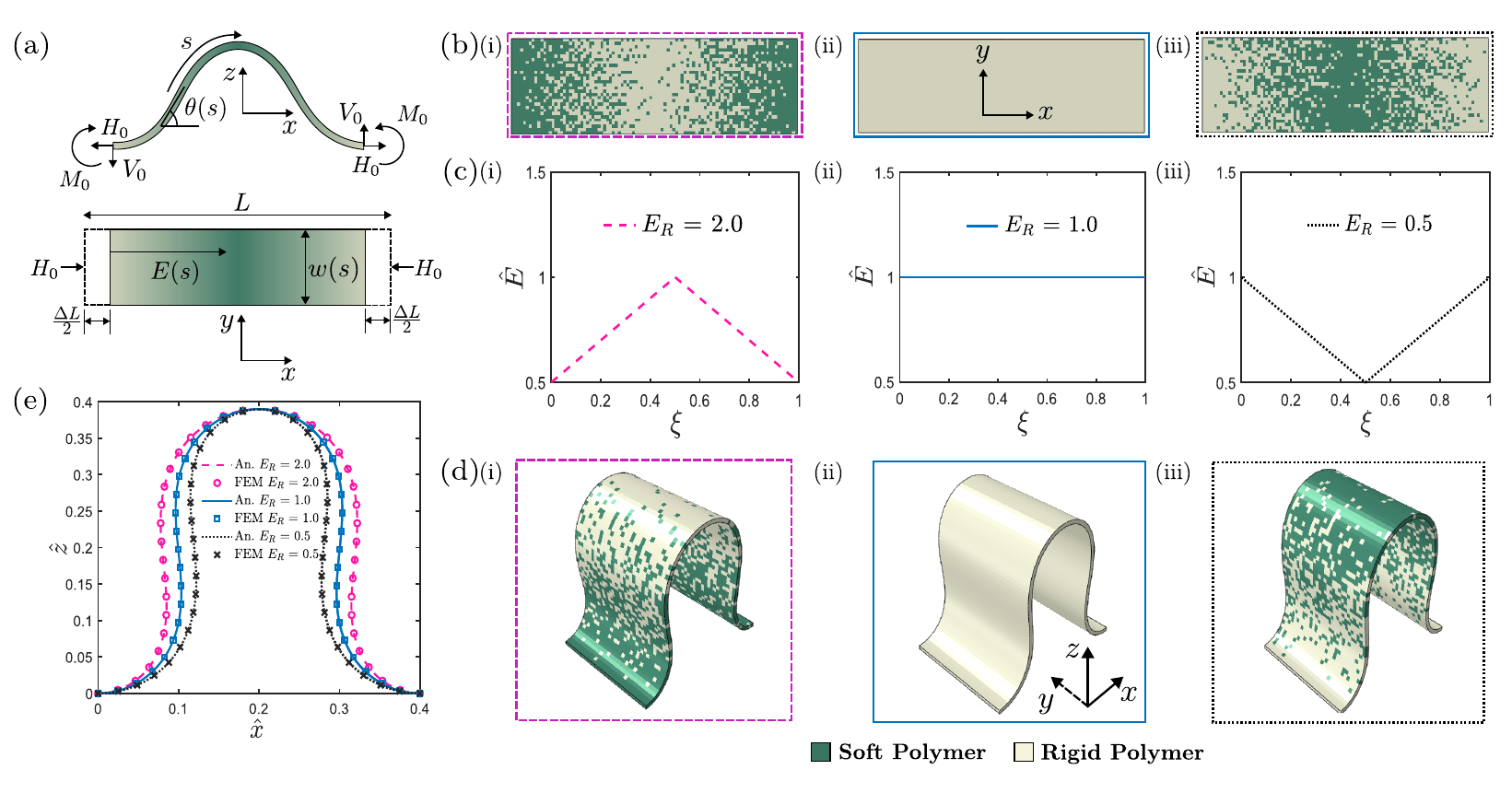} \caption{\label{fig:graded rec}Bending deformation of rectangular FGC strips. (a) Illustration of a graded strip with distributed local modulus $E(s)$ subjected to horizontal ($H_0$) and vertical ($V_0$) forces on either edge. The deformed shape is described by the equation $\theta(s)$. (b) Depicts three undeformed rectangular strips of uniform width ($w_0$) and thickness ($t_0$); the modulus grading across each strip is as follows: (i) $E_R = 2.0$, (ii) $E_R = 1.0$ and (iii) $E_R = 0.5$, where $E_R=\hat{E}({\xi=0.5})/\hat{E}({\xi=0})$. (c) The plot of the linearly varying modulus distribution across the length of each rectangular strip. (d) Deformed shapes of the FGC strips obtained from FEA simulations subjected to compressive loads applied at each edge of $\Delta L/L = 0.6$, with clamped boundary conditions. (e) Dimensionless profiles for deformed FGC strips, in which  lines indicate results predicted by analytical (An.) solution and symbols correspond to the shapes obtained from FEM simulations.}
   
\end{figure}

Considering an elastic rectangular strip of uniform thickness and width that is subjected to horizontal and vertical forces ($H_0$ and $V_0$), as well as moment ($M_0$), at both ends as depicted in Fig.~\ref{fig:graded rec}(a). Together with the bending stiffness equation obtained above, the tapered elastica equation \cite{liu2020tapered,  stuart2000buckling} that satisfies the intrinsic equation, $F[\theta(s),s)] = 0$, where $\theta(s)$ is the function describing the buckled shape, for a strip is written as

\begin{equation}
    -\frac{d}{ds}\left [B(s) \frac{d\theta(s)}{ds} \right ] = w_0[(H_0\sin\theta(s)+V_0 \cos \theta(s)].
    \label{eq:tapered elastica equation}
\end{equation}

The shape of the deformed elastica in terms of coordinates $x(s)$ and $z(s)$ can be obtained from the shape function, $\theta(s)$, by solving the geometric relationships, i.e.~$dx/ds = \cos [\theta(s)]$ and $dz/ds = \sin [\theta(s)]$. It should be noted that solving the problem requires at least one of the reaction forces to be non-zero.

Eq.~(\ref{eq:tapered elastica equation}) can be solved by applying appropriate boundary conditions (for example, the clamped boundary at both edges) together with the geometric constraint condition (i.e.~the beam accommodating a fixed amount of compression, $\Delta L$)
\begin{equation}
\left\{
    \begin{array}{lr}
    \theta(0) = \theta(L) = 0 &  \\
    \int^L_0 \cos\theta \, ds = L-\Delta L = L(1-\bar{\Delta}), & 
    \end{array}
\right.
\end{equation}
where $\bar{\Delta} = \Delta L/ L$.

Non-dimensionalisation is utilised to scale the coordinates by the length of the elastic strip, $L$, i.e.~$(\xi, \hat{x}, \hat{y}, \hat{z}) = (s, x, y, z)/L$. Similarly, the width and Young's modulus are normalised by width and the modulus at the starting position ($s = 0$) of the elastic strip, i.e.~$\hat{w}(\xi) = w(s)/w_0, \; \hat{E}(\xi) = E(s)/E_r$, where $w_0 = w(0)$, and $E_r$ is the Young's modulus of the rigid phase in FGCs, which will be discussed in details later. Accordingly, Eq.~(\ref{eq:tapered elastica equation}) becomes

\begin{equation}
    \frac{d}{d\xi}\left [ \hat{E}(\xi) \hat{w}(\xi)  \frac{d\theta}{d\xi} \right ] = -\hat{H}\frac{d\hat{z}}{d\xi}-\hat{V}\frac{d\hat{x}}{d\xi},
    \label{eq: graded tapered elastica equation}
\end{equation}
where $\hat{H}$ and $\hat{V}$ are the non-dimensional forces with the expressions, $\hat{H} = (w_0L^2/E_0I_0) H_0$ and $\hat{V} = (w_0L^2/E_0I_0) V_0$. According to Eq.~(\ref{eq: graded tapered elastica equation}), the deformed shape of a graded elastica with given geometric and material parameters can be fully determined.

The above equation can be referred to as the `Graded Elastica' equation. Nonetheless, the main difference between the Eq.~(\ref{eq: graded tapered elastica equation}) and Eq.~(8) presented in \cite{liu2020tapered} is the varying Young's modulus $\hat{E}(\xi)$ that replaces the varying thickness term $\hat{T}(\xi)^3$, where $\hat{T}(\xi) = t(\xi)/t_0$. We propose to achieve the graded modulus profile by using composite materials. The modulus of the graded elastica can be varied by combining two or more constituent materials at the localised area, following the rule obtained from micro-mechanics model. The details will be explained in Section \ref{section:composite micromechanics}. Here it should be noted that the proposed framework allows the combination of a wide range of materials as long as good interface cohesion between them is ensured during the fabrication. 

\subsection{Micro-mechanics of voxelated composites}
\label{section:composite micromechanics}

Graded composites consist of two or more phases, whereby the gradient distribution of modulus can be achieved by spatially varying the volume fraction of the embedded phases. These relationships describing the distribution of Young's modulus can readily be implemented in FEM simulations. However, it is essential to recognise how modulus gradient can be achieved during fabrication. Consequently, a manufacturing-centric approach has been adopted here for designing the FGC-based shape-morphing structures --- one strategy involves the discretisation of 3D geometry using voxels. This approach allows structures to be accurately and reliably reproduced \cite{mirzaali2020mechanics}, without relying on the generation of complex patterns to achieve modulus grading. However, a rational basis for determining the composition of each material is required to achieve the necessary modulus profile.

\begin{figure}[!ht]
  \centering
   \includegraphics[width=1\textwidth]{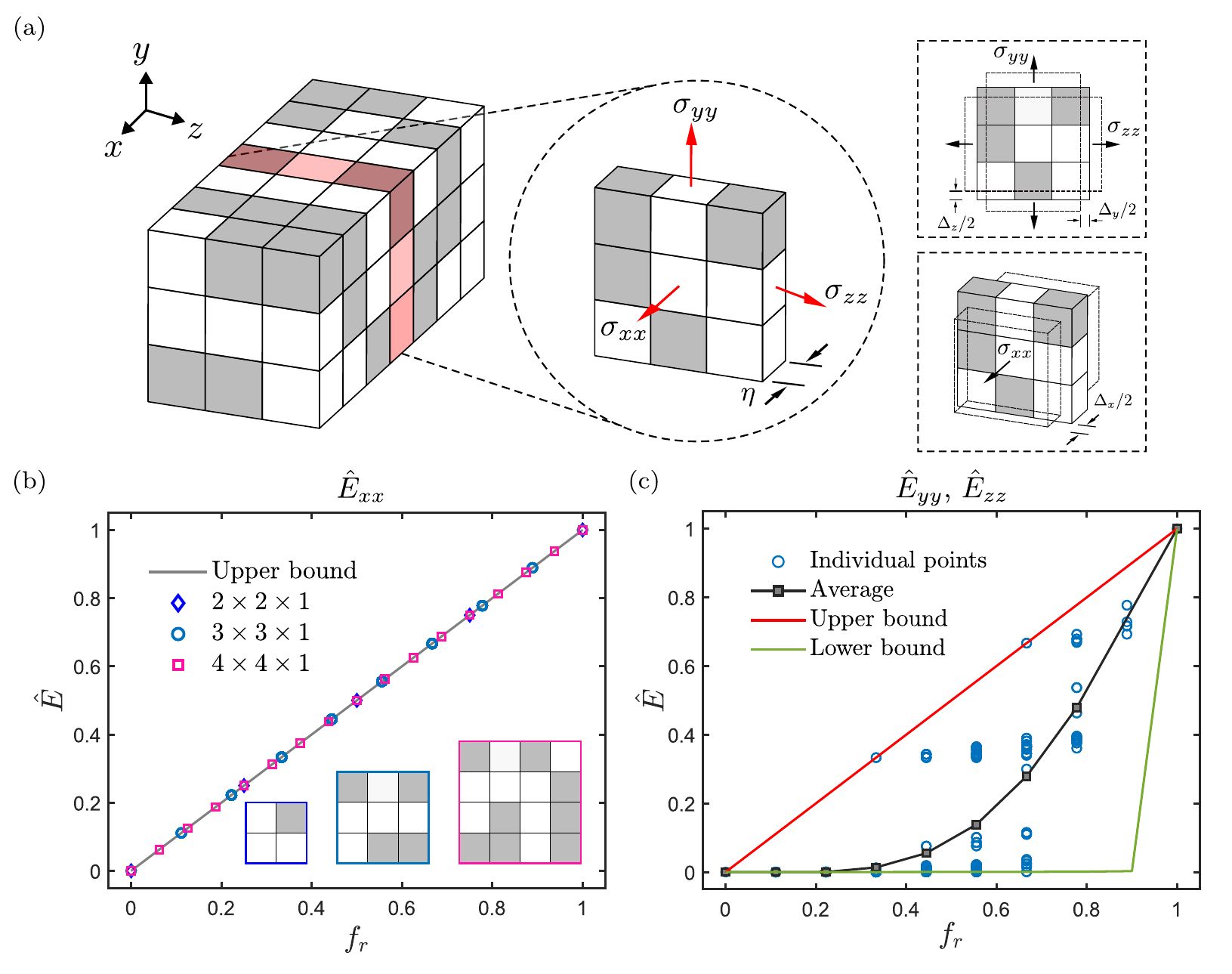}
   \caption{\label{fig:rule of mixtures}Characterisation of the Young's Moduli of FGCs with voxelated geometries. (a) Depicts a voxelated FGC block with randomly distributed rigid and soft materials. A cross-sectional one-voxel thick slice is taken to serve as a unit cell. Tensile loads of equal magnitude are applied onto either surface of the unit cell in each direction, and the corresponding moduli are measured. Numerical results of composite moduli obtained from simplified unit cells in (b) the Longitudinal direction, $\hat{E}_{xx}$, (c) the Transverse, $\hat{E}_{yy}$, and the Out-of-Plane, $\hat{E}_{zz}$, directions, as a function of the volume fraction of rigid material, $f_r$. Lower and upper bounds are obtained from Eq.~(\ref{equ: voigt bound}) and Eq.~(\ref{equ: reuss bound}). Note that the Young's modulus of soft and rigid materials used in this analysis are $E_s = 0.62$ MPa and $E_r = 2033.33$ MPa, respectively.}
\end{figure}

We start by considering a cuboid that is discretised using voxels as illustrated in Fig.~\ref{fig:rule of mixtures}(a). The discrete cuboid is composed of two materials, one of which is soft (with Young's modulus $E_s$) and the other is rigid (with modulus $E_r$), that are randomly distributed. The modulus distribution of the composite is dependent on the volume fraction and on the position of each material within the geometry. Consequently, a range of moduli can be attained, which expresses the requirement to explore all possible combinations. This is to determine all arrangements of voxels and their overall effects on the composite modulus. The relationship between the material's position within given geometry and its effect on the modulus could be determined using machine learning algorithms such as artificial neural networks. However, this would necessitate the collection of sufficient data for training such that accurate predictions can be made, which could in turn be resource intensive. As a consequence, from the micro-mechanics point of view, probing a unit cell would be a feasible alternative.

Again, considering the discretised cuboid shown in Fig.~\ref{fig:rule of mixtures}(a), the unit cell is of dimensions $3 \times 3 \times 1$ voxels. This would mean the number of material configurations within the unit cell is reduced to $2^9 = 512$. Furthermore, stresses in the principal directions were applied onto the surface of the geometry, denoted to be $\sigma_{ii} ~(i=x,y,z)$. This is to obtain the modulus in the longitudinal ($x$), transverse ($y$), and out-of-plane ($z$) directions. The resulting moduli for all 512 cases are plotted against the volume fraction of the rigid material, as shown in Fig.~\ref{fig:rule of mixtures}(b)-(c). Interestingly, the longitudinal modulus,  $\hat{E}_{xx}$, scales proportionally with the volume fraction of rigid material, $f_r$, and by extension, the volume fraction of soft material $f_s$ ($ = 1-f_r$). This means that regardless of the configuration of the materials in the unit cell, a unique longitudinal modulus is obtained, that scales with the volume fraction. This occurs due to the affine deformation across the voxel-thick unit cell of size $\eta$, whereby an applied global deformation $\varepsilon_{xx}$ is translated uniformly to microscale \cite{flory1982theory}. It should be noted that $\eta \ll L$, where $L$ is the length of the structure and $\Delta_x$ is the change in length. Therefore, the strain on soft phase $\varepsilon_s$ is equal to the strain across the rigid phase $\varepsilon_r$.

\begin{equation}
    \varepsilon_{xx}=\frac{\Delta_x}{\eta} = \varepsilon_s = \varepsilon_r.
    \label{equ: affine strain}
\end{equation}

Upon plotting the outcome, we realise that the results simply follow the Voigt bound of the rule of mixtures \cite{voigt1889}. Consequently, the following equation can be used to calculate the modulus in the longitudinal direction $\hat{E}_{xx}$,

\begin{equation}
    \hat{E}_{xx}(\xi) = \left [f_r(\xi){E}_r+ f_s(\xi){E}_s\right ]/E_r,
    \label{equ: voigt bound}
\end{equation}

Since the bending deformation is primarily dependent on the longitudinal modulus, we only need to match the modulus profile in the longitudinal ($x$) direction. This allows the modulus to be varied at position $\xi$, along the length of the elastic strip. Further FEM simulations have also been conducted on geometry with a reduced number of voxels $2 \times 2 \times 1$ as well as geometry with a greater number $4 \times 4 \times 1$. It was found that the longitudinal modulus is independent of the number of voxels in the $y-z$ plane as long as the voxel-thick block experiences a small strain (see Fig.~\ref{fig:rule of mixtures}(b)). This can be ensured by utilising a beam of sufficiently large aspect ratio (Here aspect ratio is defined as length over the thickness).

Alternatively, plotting the moduli obtained from the transverse ($y$) and out-of-plane ($z$) directions ($\hat{E}_{yy}$ and $\hat{E}_{zz}$), as shown in Fig.~\ref{fig:rule of mixtures}(c), there exist multiple solutions for the same volume fraction, $f_r$. In both $y$ and $z$ directions, the cuboid does not undergo the same amount of deformation subjected to constant stress, therefore, would result in vastly differing composite moduli. This results from the mismatch between the local strains of constituent materials and the applied global transverse strain, so-called ``non-affine deformation" \cite{miehe2004micro}. The combination of moduli obtained here would lie between the Voigt and the Reuss bound, where the Reuss bound is obtained using the following equation \cite{reuss1929},

 \begin{equation}
    \hat{E}_c(\xi) = \left[ f_r(\xi) + \frac{f_s(\xi)}{(E_s/E_r)} \right]^{-1}.
    \label{equ: reuss bound}
\end{equation}

Nevertheless, there also exist other sophisticated models that may capture the micro-mechanics of such multi-phase materials, such as the Hashin and Shtrikman \cite{hashin1963variational} with compact upper and lower bounds but at a cost of heightened complexity.

\subsection{Bending deformation of a rectangular FGC strip}
\label{section:benchmabenchmark study}

According to Eq.~(\ref{eq:tapered elastica equation}), by changing the modulus distribution, deformed shapes with different curvatures can be achieved. The FEM simulation was employed to examine the effect of varying modulus on the deformation of a rectangular strip. As shown in Fig.~\ref{fig:graded rec}(a), three rectangular strips of the same dimensions subjected to horizontal forces at both edges, of equal magnitude, were considered. The modulus distribution was linearly varied, as shown in Fig.~\ref{fig:graded rec}(b) and (c): from (i) to (iii) corresponding to the modulus ratio $E_R=\hat{E}({\xi=0.5})/\hat{E}({\xi=0})=$ 2.0, 1.0, and 0.5, respectively. Noted that the green and white voxels correspond to the soft and rigid materials, respectively. Fig.~\ref{fig:graded rec}(b \& c)-(i) represents a linear increase in the stiffness of the geometry as it reaches the centre of the geometry, followed by a linear decrease. In contrast, Fig.~\ref{fig:graded rec}(b \& c)-(iii) shows a linear reduction in the stiffness, having the lowest stiffness at either end and reaching a peak of $\hat{E} = 0.5$. Finally, the rectangular strip in Fig.~\ref{fig:graded rec}(b \& c)-(ii) follows a rectangular strip with a uniform modulus, made from a rigid polymer. Accordingly, the deformed shapes of those three strips obtained in FEM simulations are displayed in Fig.~\ref{fig:graded rec}(d).

A comparison between the analytical predictions and FEM simulations of the deformed shapes is presented in Fig.~\ref{fig:graded rec}(e), where the types of lines and colours correspond to the shapes in Fig.~\ref{fig:graded rec}(d). A good agreement between the theoretical predictions and FEM simulations is achieved. As anticipated, a change in local modulus affects the curvature of the resultant shape when deformed. Following the dashed pink line, by increasing the modulus at the middle part of the rectangular strip, the buckled shape obtained is relatively wider in comparison to the strip with a uniform modulus. However, with a reduction in modulus at the middle part, as shown by the dotted black line, the buckled shape is narrower than the uniform strip. Intuitively, by increasing the local modulus, it would be equally difficult to bend that portion of the strip as a result of increased local bending stiffness, resulting in reduced curvature of the deformed shape. Following these results, the inverse design of a 3D structure can be addressed. Such shapes can be obtained through the selection of appropriate local bending stiffness, achieved by tapering each strip and controlling the local modulus.

\section{Inverse design of 3D tessellated morphing structures based on FGCs}
\label{section:inverse design}

\subsection{Design principle}
\label{section:design principle}

\begin{figure}[ht]
  \centering
   \includegraphics[width=1\textwidth]{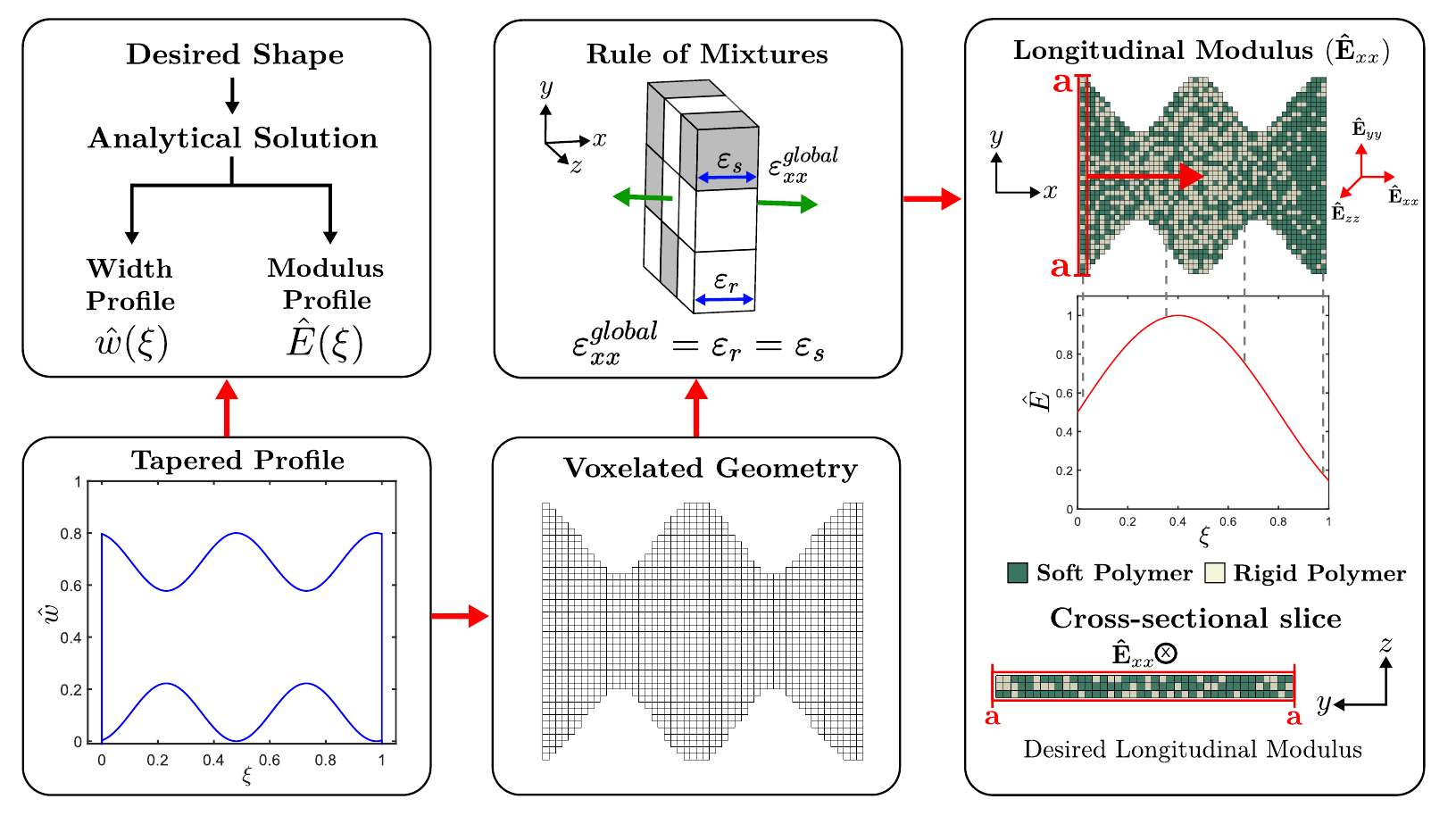}
   \caption{\label{fig:design principle}Illustration of the design strategy employed to obtain a structure with varying Young's modulus. The desired shape is defined through a local inclination angle ($\theta({\xi})$), that relates the arc length ($s$) to the corresponding arc angle. Substituting this equation into the analytical solution Eq.~(\ref{eq: graded tapered elastica equation}) to obtain the tapering and modulus distribution. The geometry is then discretised using voxels of size $\eta$. By considering cross-sectional slices of geometry, the volume fraction required to reach the required Young's modulus can be calculated using the rule of mixtures. This allows the longitudinal modulus to follow the exact modulus-grading profile.}
\end{figure}

To achieve a desired 3D shape, the graded elastica equation presented in Section \ref{section:theory rectangular strip} can be employed to obtain the width profile, $\hat{w}(\xi)$, and the modulus profile, $\hat{E}(\xi)$; and the voxel-based 3D printing can be harnessed for fabrication. On these bases, a regularised design framework is proposed in Fig.~\ref{fig:design principle}. In essence, the width profile describes the cut pattern of the flat 2D sheet, while the modulus profile dictates the local curvature of the 3D morphed structure. The geometry of the 2D sheet as detailed by the width profile is discretised into voxels of size $\eta$ thus simplifying the process of assigning material to each voxel. By considering one-voxel-wide cross-sectional slices in the $y-z$ plane, along the $x$ axis, the distribution of each material in the slices can be calculated. To control the longitudinal modulus ($E_{xx}$) of each slice, the rule of mixtures based on Voigt bound (equal strain assumption $\varepsilon_{x}$, as discussed in Section \ref{section:composite micromechanics}, see Eq.~(\ref{equ: voigt bound})) can be applied to find the volume fraction of both rigid and soft materials. The modulus profile from the analytical solution (it will be discussed in Section \ref{section:theoreticalformulation}) serves as the modulus of the FGCs, normalised by the modulus of rigid material. Nonetheless, it is necessary to ensure that the mixture of soft and rigid polymers is able to cover the range of modulus attained from the analytical solution, i.e. the modulus of soft material must be less than or equal to the minimum modulus along the elastica, $E_s/E_r \leq \min\{E({\xi})\}$. Consequently, the number of voxels ($N_t$) containing either material is obtained by calculating the volume fraction required to achieve the modulus of the composite. This is achieved using the Voigt bound of the rule of mixtures (Eq.~(\ref{equ: voigt bound})), with the number of voxels containing the soft $N_s$, or rigid $N_r$, material is calculated using the following,

\begin{equation}
    N_t = \frac{\hat{w}(\xi)\hat{T}(\xi)}{\eta}, \qquad  N_r = N_t  V_r, \qquad N_s = N_t - N_r.
\end{equation}

This process is repeated along the length of the elastic strip, thus allowing the composite to be printed with varying longitudinal modulus. Following the discussion in Section \ref{section:composite micromechanics}, it should be emphasised that only the longitudinal modulus ($E_{xx}$) here is controlled, the resulting transverse and out-of-plane moduli ($E_{yy}$ and $E_{zz}$) are random due to non-uniform strain across the rigid and soft materials in the those directions.

\subsection{Theoretical formulation}
\label{section:theoreticalformulation}

Distinctly, 3D tessellated structures can be formed by connecting several tapered (or more precisely, modulus-graded) elastic strips, i.e.~`spokes', to a central hub \cite{liu2020tapered}. This work focuses on two parameters, namely the width profile, $\hat{w}(\xi)$, and the modulus distribution, $\hat{E}(\xi)$. The resulting expressions satisfy the tessellation condition as well as match the required local bending stiffness for a 3D target shape with given curvature distribution.

By integrating Eq.~(\ref{eq: graded tapered elastica equation}) and rearranging, an explicit equation can be obtained as

\begin{equation}
    \hat{E}(\xi) \hat{w}(\xi) = \frac{\hat{H}(\hat{z}_*-\hat{z})+\hat{V}(\hat{x}_*-\hat{x})}{d\theta(\xi)/d\xi},
    \label{eq:int}
\end{equation}
where $\hat{z}_*$ and $\hat{x}_*$ are integration constants, together with $\hat{H}$ and $\hat{V}$ as unknown parameters, and their determination will be discussed later. Note that this equation explicitly relates the geometric ($\hat{w}(\xi)$) and physical ($\hat{E}(\xi)$) properties of the 2D cut pattern to the curvature information ($d\theta(\xi)/d\xi$) of the 3D tessellated target shape.

For the tessellation to be achieved, the edges of adjacent spokes must touch along the length of the spoke. Consequently, at every arc length, $\xi$, the width profile, $\hat{w}(\xi)$, must satisfy the geometric relation, $\hat{w}(\xi) = ({2L}/{w_0}) \hat{x}(\xi) \tan ({\pi}/{N})$, where $N$ is the number of spokes to be chosen, {$\hat{x}(\xi) = 1 + \hat{r} - \bar{\Delta} - \int^\xi_0 \cos \theta(\xi ')d\xi'$ is used to determine the $x$-coordinate of a particular deformed point, and $\hat{r}$ is the normalized radius of the central hub.}

The equation relating the width profile, $\hat{w}(\xi)$, and the local inclination angle, $\hat{\theta}(\xi)$, can be found using the following relation
{
\begin{equation}
    \hat{w}[\xi; \hat{\theta}(\xi)] = \frac{2L}{w_0} \cdot \tan \left(\frac{\pi}{N} \right) [\hat{r} + \hat{X}(\xi)],
    \label{eq: tessellation condition}
\end{equation}}
where $\hat{X}(\xi) = 1 - \bar{\Delta} - \int^\xi_0 \cos \theta (\xi ')d\xi'$.

Substituting Eq.~(\ref{eq: tessellation condition}) into  {Eq.~(\ref{eq:int})} and applying appropriate boundary conditions, 3D axisymmetric structures can be realised. However, by simply following the tessellation condition but keeping the modulus profile as uniform, the desired tessellated structures will not be achieved. Therefore, to match the desired shape, the local modulus, $\hat{E}(\xi)$, needs to be related to the curvature of the target structure, $\theta (\xi)$, as 

\begin{equation}
    \hat{E}(\xi) = \frac{\hat{H}(\hat{z}_*-\hat{z})+\hat{V}(\hat{x}_*-\hat{x})}{\hat{w}(\xi) \theta_{\xi}}.
    \label{eq:1}
\end{equation}

If we only consider the horizontal force, the above equation can be simplified as 

\begin{equation}
    \hat{E}(\xi) = \frac{\hat{H}(\hat{z}_*-\hat{z})}{\hat{w}(\xi) \theta_{\xi}}.
    \label{eq:no hoz force}
\end{equation}

Ultimately, by utilising the curvature distribution of a target 3D shape, the width profile of individual spoke and the corresponding modulus distribution can be obtained by solving Eq.~(\ref{eq: tessellation condition}) and Eq.~(\ref{eq:1}) (or \ref{eq:no hoz force}), respectively.

However, the unknown parameters $H_0$ and $z_*$ must also be determined by solving the inverse problem. Following the approach developed by Liu et al.~\cite{liu2020tapered}, these parameters can be obtained by considering two cases, based on whether an inflection point exists or not in accordance with appropriate boundary conditions. The desired 3D shape is not limited to positive or negative global Gaussian curvature, varying Gaussian curvature can also be achieved. The solutions of unknown parameters in case \textbf{(a) no inflection point} in the desired profile and case \textbf{(b) with inflection point} can be obtained as follows,

\begin{subequations}
\begin{align}
    & \hat{H} = \frac{w(0)\theta_\xi(0)-w(1)\theta_\xi(1)}{y(1)}, \; z_* = \frac{w(0)\theta_\xi(0)}{\hat{H}}, 
    \label{eq:no inflection}
    \\
    & z_* = \int_0^{\xi_{*}} \sin\theta d\xi, \; \hat{H} = \frac{\theta_\xi(0)}{z_*}.
    \label{eq:inflection}
\end{align}
\end{subequations}

Noted that, in Eq.~(\ref{eq:inflection}), $\xi = \xi_{*}$ is the position of the inflection point, i.e. $\theta_\xi(\xi_{*})=0$. By substituting the solutions from Eq.~(\ref{eq:no inflection}) or Eq.~(\ref{eq:inflection}) into Eq.~(\ref{eq:no hoz force}), in conjunction with the tessellation condition described by Eq.~(\ref{eq: tessellation condition}), the inverse design problem can be fully addressed. The same procedure can be applied if vertical forces need to be considered.

\subsection{Multi-material additive manufacturing}
\label{section:printing process}

Ensuring the compatibility between two polymers is a critical factor in the fabrication and maintenance of fundamental functionality of FGC-based shape-morphing structures. There are several techniques that can be used to achieve the compatibility between two different materials joined by interface, including but not limited direct joining, gradient path and intermediate section \cite{yan2020additive} (see Fig.~\ref{fig:combining two polymers} in \ref{section:joining polymers} for more details). These techniques facilitate the creation of compliant materials that can bend without failure. Nonetheless, the choice of technique used to combine the two materials depends on the specific material selected. In this work, to fabricate the proposed FGC-based shape-morphing structure, a combination of direct joining and gradient path is utilised.

\begin{figure}[!h] 
   \centering 
   \includegraphics[width=1\textwidth]{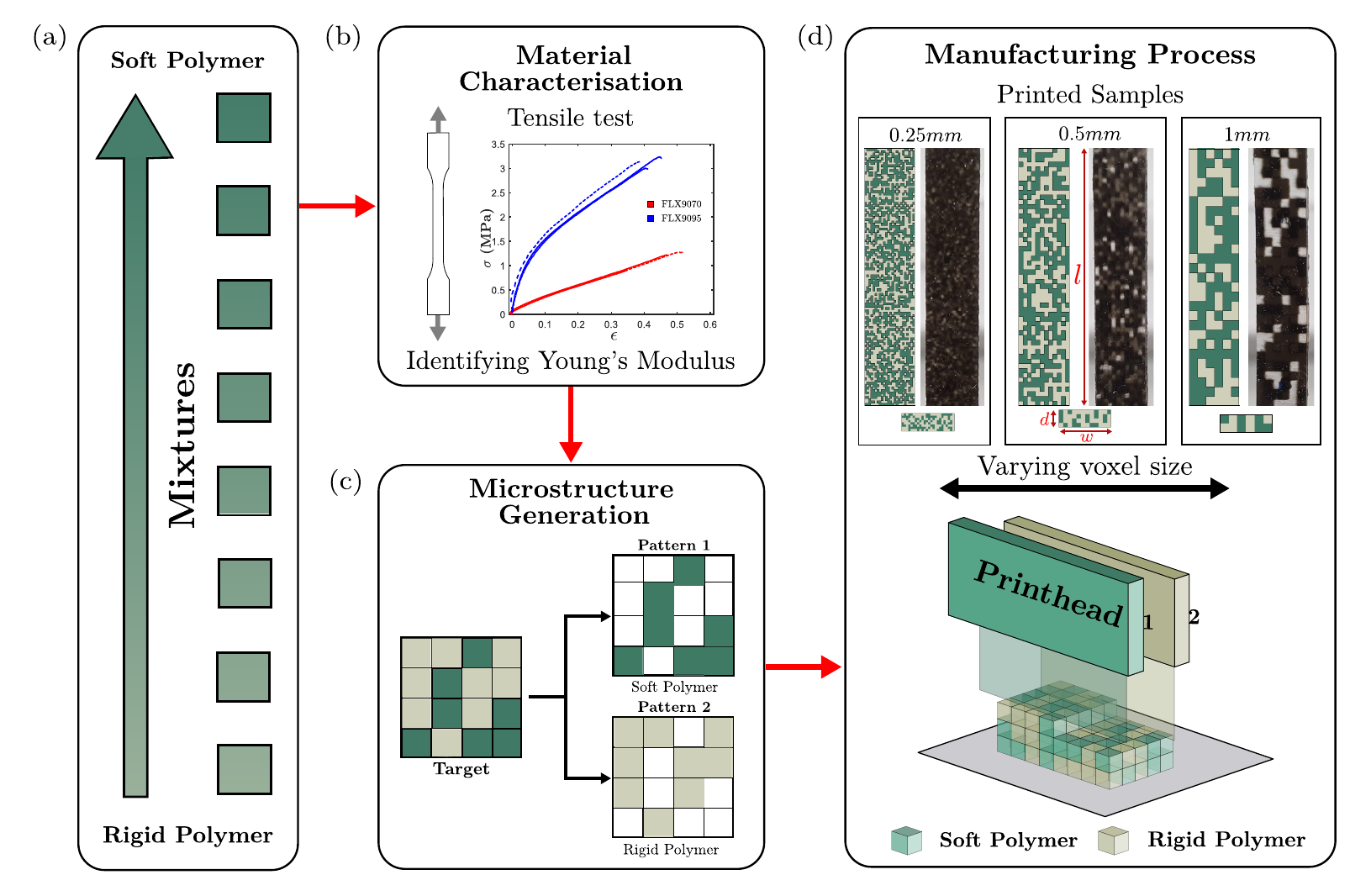}
   \caption{\label{fig:printing process}Steps illustrating the printing process, from the material selection to generating files and an example of the final product. (a) Soft and rigid polymer resins were blended at different proportions to produce composite mixtures. (b) Tensile tests following ISO-527-1 were undertaken to characterise material properties for each polymer blend and Young's modulus was extracted. The graph shows various stress-strain curves for two polymers (FLX9070 and FLX9095). (c) Two CAD files were generated using ABAQUS, each corresponding to one of the two polymers. (d) The CAD files were then used in the additive manufacturing process using Stratasys Objet260 Connex. Some 3D printed samples with different voxel sizes are shown with the corresponding FEM file of dimensions $l = 35$ mm, $w = 6$ mm, and $d = 2$ mm.}
\end{figure}

Fig.~\ref{fig:printing process} shows the steps involved in the manufacturing of the FGCs, used to form a 3D tessellated shape-morphing structure. The FGC strips were produced using Stratasys Objet260 Connex, a material jetting technique utilising photoreactive polymer resins. This form of additive manufacturing process is comprised of a series of alternative steps. The printing process involves jetting a layer of photopolymer particulates from each printhead, and these layers are then UV cured \emph{in situ} as the printhead moves along the x-axis of the build plate, with the build plate lowering after successive material deposition. Prior to the material deposition, the composite resin is heated to achieve the desired viscosity, which further strengthens the bonding between the two materials within the composite resin. The multi-material printing process primarily utilises two different resins. 

There are two types of resin commonly used, the first type is the individual photopolymers, such as the soft and rubbery resin, the so-called ``TangoBlack", and the rigid and glassy one, ``VeroClear". They can be directly used for two-material printing, however, their moduli are significantly different and not easy to be combined to map a range of modulus' distribution. The alternative type is the pre-mixed resins by combining ``TangoBlack" and ``VeroClear" of various volume fractions. We characterised these resins via tensile tests, and the measured Young's moduli are summarised in Fig.~\ref{fig:resin modulus} in \ref{section:resin modulus}. To match the requirement of the modulus' ratio in the two-material printing, we finally chose two pre-mixed resins ``FLX9070" and ``FLX9095" (the corresponding stress-strain curves under tensile tests are shown in Fig.~\ref{fig:printing process}(b)) for printing the samples in the following experiments.

By integrating multiple printheads (each taking charge of one resin), the machine allows the user to co-print multiple resins in order to create polymer composites with tunable mechanical properties. In order to programme the mechanical response of the printed FGC sample, we propose the voxel-based design strategy, as shown in Fig.~\ref{fig:printing process}(c): the user can decide the size, material and location of each voxel, stack them up to construct the final geometry with desired performance. Using ABAQUS python scripting, two complementary CAD files for 3D printing are generated. Both of these files outline the pattern and distribution of voxels required to be printed, and each corresponds to one of the chosen materials. The aggregate of these two complementary patterns forms the target geometry with the required material compositions, and therefore, the desired distribution of mechanical properties.

Fig.~\ref{fig:printing process}(d) shows some printed FGC samples with dimensions $l = 35$ mm, $w = 6$ mm, and $d = 2$ mm for mechanical testing. The overall composition of the structure was designed to contain $50\%$ of voxels in TangoBlack and $50\%$ in VeroClear, with random distribution across the geometry. Three different voxel sizes of $\eta = 0.25$ mm, 0.5 mm and 1.0 mm were manufactured. Considering the geometric discretisation and the voxel-based printing are time-consuming, printing the samples with different voxel size was done to determine the printing resolution achievable by the printer, with a trade-off between the resolution and the time taken to generate and print the discretised geometry. It should be noted that the voxel size also affects how well the FGC strip follows the modulus profile. Finally, we fixed the voxel size as $\eta = 0.5$ mm for all the FGC samples in the 3D printing process.

\subsection{Validating 3D tessellated morphing structures}
\label{section:3DTessellation}

\subsubsection{Experimental implementation}

\begin{figure}[!h]
   \centering
   \includegraphics[width=1\textwidth]{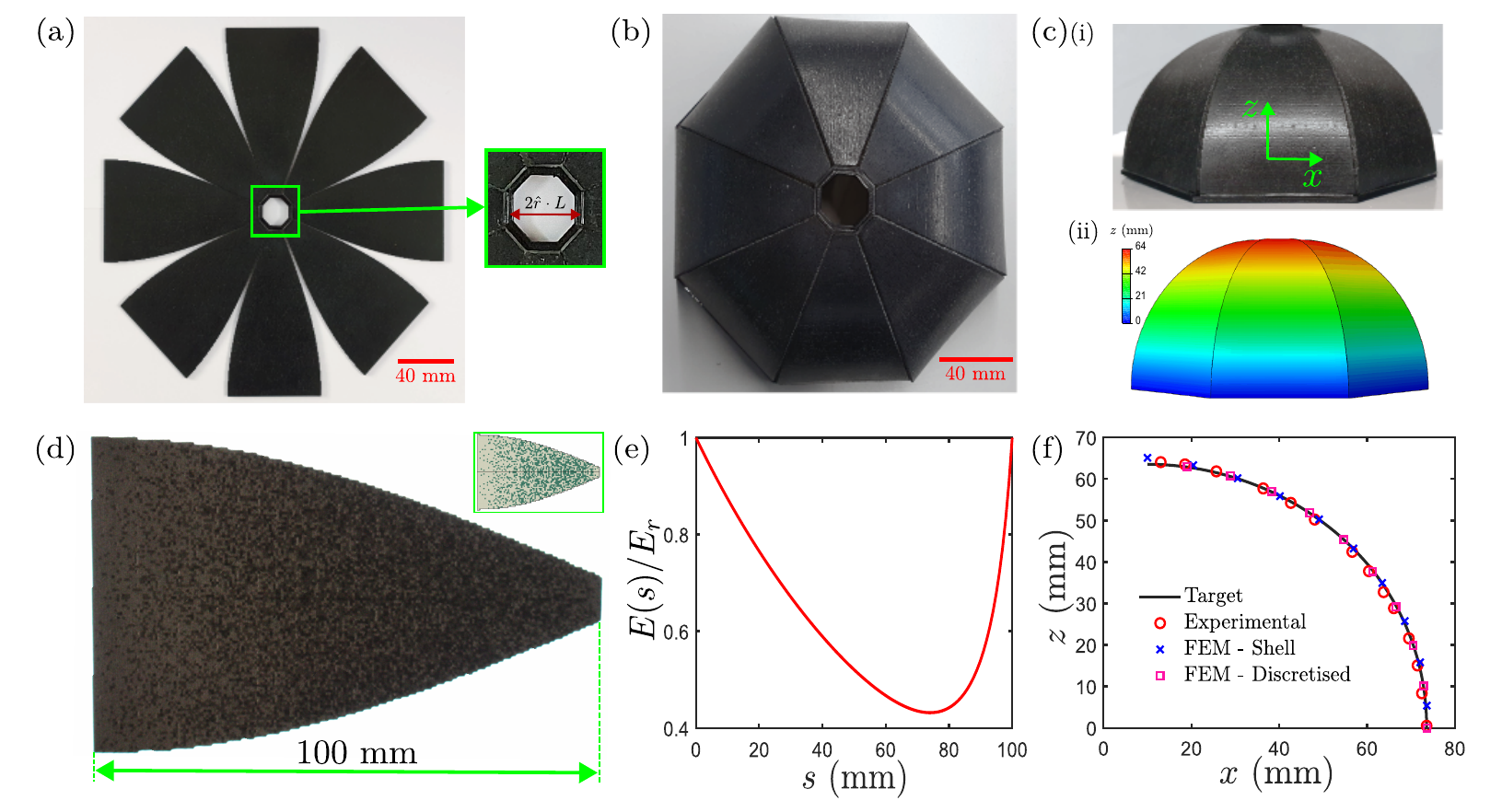}
   \caption{\label{fig:tessellated hemisphere}Demonstration of the inverse design of 3D tessellated hemisphere. (a) Printed 2D pattern with tapering and modulus grading described by the analytical solution to form 3D hemisphere. (b) Top view of the morphed hemisphere with tessellation. (c) Side views of the morphed hemisphere: (i) fully assembled 3D printed structure; (ii) Deformed profile obtained from FEM simulation using shell-based smooth width profile. (d) Depicts the grading across a single FGC strip (with thickness $t = 2$ mm and width profile following $\hat{w}(\xi)  100$ mm with voxel size $\eta = 0.5$ mm) obtained from 3D printing, top right showing same geometry generated using FEM. (e) The normalised modulus distribution along the strip following the design principle from Fig.~\ref{fig:design principle}. (f) Comparison between the target and the deformed 3D profiles obtained from both the experiment and FEM simulations from shell-based model and discretised-solid element model.}
\end{figure}

To demonstrate the inverse design strategy, a hemisphere was chosen as a simple example of the target 3D shape. Following the procedure described in Section \ref{section:design principle}, the 2D cut pattern of the plate made of FGC with width profile $\hat{w}(\xi)$ and modulus profile $\hat{E}(\xi)$, as shown in Fig.~\ref{fig:tessellated hemisphere}(a), are obtained according to the analytical solutions, specifically combining Eqs.~(\ref{eq: tessellation condition}) and (\ref{eq:no hoz force}) with Eq.~(\ref{eq:no inflection}) as there exists no inflection point within the profile of the chosen target shape. By applying a horizontal force onto the outer edge of each FGC strip, the 2D flat plate can be morphed into a 3D hemisphere, as shown in Figs.~\ref{fig:tessellated hemisphere}(b) and (c). Eight strips ($N=8$) were chosen to form the hemisphere, with the inner radius chosen to be $R = 0.1$. As shown in Fig.~\ref{fig:tessellated hemisphere}(d), each strip is of length $L = 100$ mm and thickness $t_0 = 2$ mm with voxel size $\eta = 0.5$ mm, while the width profile and the modulus' distribution (see Fig.~\ref{fig:tessellated hemisphere}(e)) follows the analytical solution. These FGC strips were manufactured using 3D printing (polymer jetting), as discussed in Section \ref{section:printing process}. FLX9070 and FLX9095 are two flexible polymers used in printing. These two materials were utilised as $E_s/E_r = 0.25$, hence it can cover the range of modulus requirement such that $E_s/E_r \leq \min\{\hat{E}(\xi)\}$ or $E_s \leq \min\{E({s})\}$, used to achieve the modulus' distribution, as shown in Fig.~\ref{fig:tessellated hemisphere}(e). In total, it took one hour and a half to print the eight FGC strips with the Objet 260 Connex. This inkjet printer has a high printing resolution of \SI{44}{\micro\metre} on the in-plane X-Y axis and \SI{16}{\micro\metre} on the through-thickness Z axis. It also can be expected that the printing resolution and printing time should be able to further enhanced with the availability of newer printer models. 


After the printing of eight FGC strips, a two-step assembly method was used to construct the complete 3D tessellated hemisphere. The first step involved connecting the strips to a hollow central hub, which was 3D printed using Polylactic Acid (PLA) with Young's modulus of $\sim$ 2 GPa. Double-sided adhesive pads were then used to join the inner edge (i.e.~the narrower side, as shown in Fig.~\ref{fig:tessellated hemisphere}(d)) of the strips to the central hub, creating a 2D `flower' pattern, as shown in Fig.~\ref{fig:tessellated hemisphere}(a). Note that it is also able to print the whole 2D `flower' pattern directly, rather than assembly after the print, by using a larger 3D printer. Next, the outer edge of each FGC strip was inserted into slot of specific width and slant angle within a base (also 3D printed using PLA), where the positions of slots are determined from the theoretical model as boundary conditions directly, forming the complete tessellated hemisphere, as presented in Figs.~\ref{fig:tessellated hemisphere}(b) and (c)-(i). 

Once we obtained the morphed 3D structure made of the printed FGCs, we can compare its profile with the target. 
Extracting the profile of the 3D morphed structure involved the use of 3D scanning camera (EinScan-SE), which utilises structured light to measure 3D shape of the object. A projector is used to create patterns of light, where the cameras then capture the distortions in the light patterns due to the object's shape. By triangulating multiple scans at various angles, the coordinates of the morphed shape can be obtained. This data can subsequently be stitched, forming a digital version of the scanned object. The digitised object file was then used to extract the coordinates of the morphed structure. Note that here the coordinate data from the morphed profile is extracted from the neutral plane of FGC strips in the thickness direction. The comparison between the measured shape of the sample in experiments (the red circles) and the shape of the target (the grey solid line) is shown in Fig.~\ref{fig:tessellated hemisphere}(f), and a very good agreement is achieved.

\subsubsection{FEM simulation}
\label{section:FEM validation}

As a further validation, we performed FEM simulations to examine the morphing process and compare the predicted profile with both the experimental profile and the target profile. Two different FEM models were employed to simulate the FGC-based morphing structure. The first one accounts for all the details of the discretised geometry, in which each voxel is explicitly assigned with associated (either rigid or soft) material and modelled by 3D solid elements. To ensure sufficient resolution, each voxel is meshed using 8 hexahedral elements, that is, of size, $\eta/2$. The second model utilises shell elements, while the tapered strip geometry (i.e.~width profile) follows the exact width distribution obtained from the analytical solution. For both two models, the predefined field is adopted to describe the distributed Young's modulus, and the equation used to describe modulus profile was fitted using particular functions (such as Gaussian model), which was then implemented as an analytical field in FEM software ABAQUS. In the simulation process, compressive (with a given displacement) and rotational (with a given angle) boundary conditions were applied to the edges of each strip. Both the soft (FLX9095) and rigid (FLX9070) phases within the FGCs were modelled as isotropic elastic solids, with Young's modulus of $E_s=$ 3.1 MPa and $E_r=$ 12.6 MPa, respectively, and both with Poisson's ratio of $\nu = 0.3$. The 3D morphed profiles obtained from these two FEM models are presented in Fig.~\ref{fig:tessellated hemisphere}(f) as squares and crosses (the profile of the morphed structure in shell-based FEM model is shown in Fig.~\ref{fig:tessellated hemisphere}(c)-(ii)). Both of them match the target shape and the profile obtained from the experiment very well. Nonetheless, it should be noted that benefiting from the axisymmetric nature of the morphed structure, only one single FGC strip needs to be modelled in the discretised FEM model to reduce the computational cost. Consequently, a visual representation of the whole FGC morphing structure can be generated using the symmetric condition, see Fig.~\ref{fig:mesh comparison}. Alternatively, in the shell-based FEM model, the whole morphing structure can be modelled at a substantially lower computational cost, further discussion can be found in Section \ref{section: load bearing FGCs}.

\section{Additional demonstrations showcasing the design of shape-morphing FGCs}

\subsection{Graded composite design through aggregate patterns}

\begin{figure}[!h] 
   \centering
   \includegraphics[width=1\textwidth]{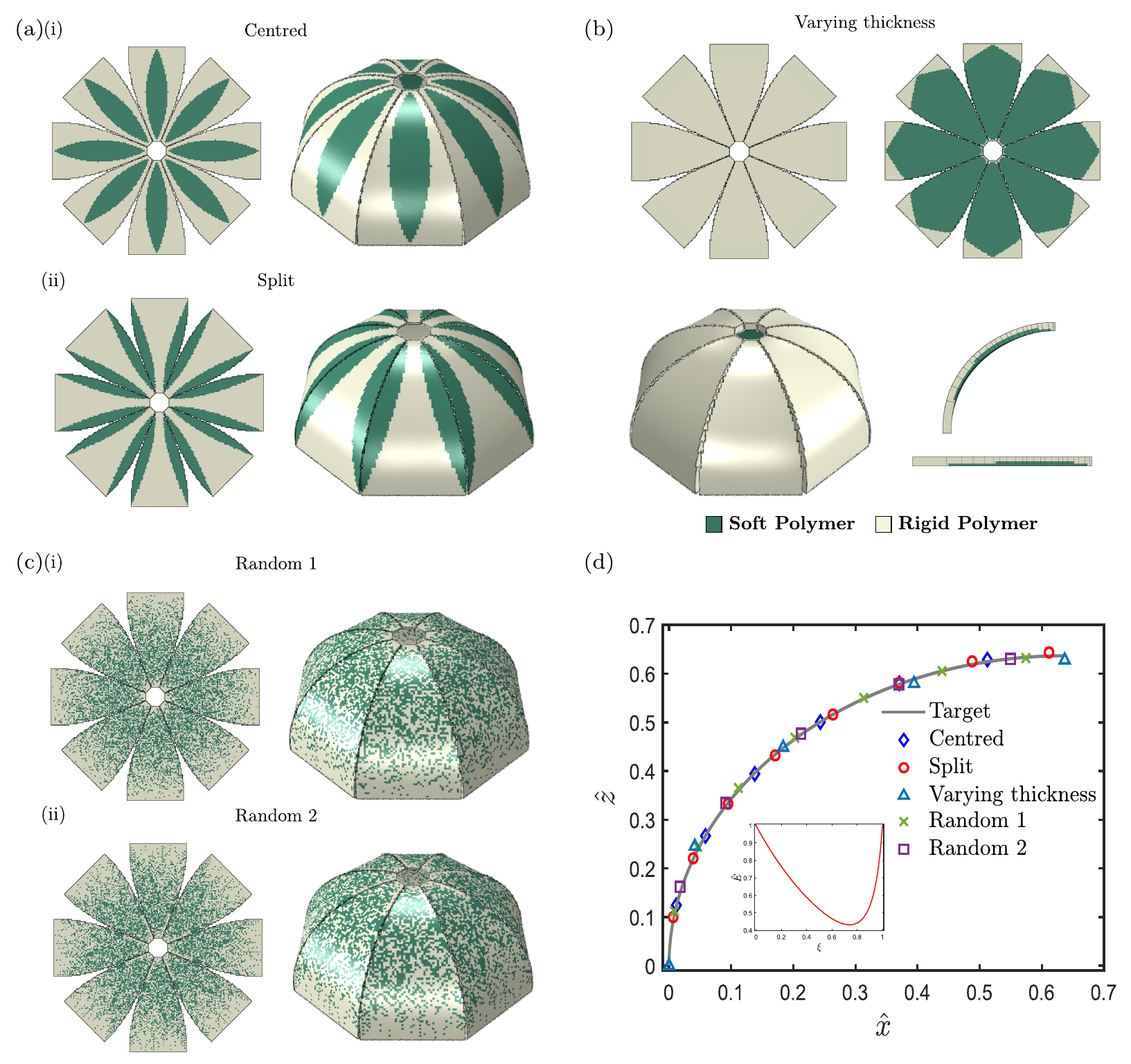}
   \caption{\label{fig:deformed_shape_d3.pdf} The FGC-based shape-morphing (hemispherical) structures with different aggregate pattern of material's distribution across (a) the width direction and (b) the thickness direction; as well as (c) random distribution. In (a), two cases are considered according the distribution of the soft polymer: (i) `Centred' and (ii) `Split'; and the initial flat state and the morphed 3D state are shown in the left- and right-hand side, respectively. In (b), the front view and the back view of the flat sheet are shown on the top; and the full view and the cross view of the morphed 3D structure are shown on the bottom. In (c), two types of random distribution of soft and rigid polymers are considered. (d) Comparison of deformation profiles obtained from FEM simulations and analytical solution with the graph in insert describing the modulus distribution across the arc-length, whereby the target shape here is a hemisphere.}
\end{figure}

We have demonstrated that the FGC-based shape-morphing structures can attain the desired shape through the utilisation of designed distribution of effective mechanical properties, but random allocation of materials, along its length. Nevertheless, we can apply the same design approach to achieve diverse patterns while adhering to the volume fraction limit imposed by the rule of mixtures. One form of pattern can be produced by grouping alike materials together, which can be formed in eithor the width direction or the thickness direction, over the length of the structure. Fig.~\ref{fig:deformed_shape_d3.pdf} illustrates the variety of patterns that can be achieved through the use of aggregate materials' distribution.

To illustrate the potential patterns that could emerge in the width direction with the target shape of a hemisphere, we begin by examining the clustering of like materials within each layer. These patterns are replicated along the thickness direction, resulting in two distinct patterns. The first pattern, referred to as `Centred' is characterised by the soft material being predominantly located in the middle of the strip. The second pattern, known as `Split' involves the total number of soft material voxels being divided into two, and concentrated along the top and bottom edges of the strip. Fig.~\ref{fig:deformed_shape_d3.pdf}(a) showcases the undeformed and deformed hemispherical shapes. It is also possible to have all the soft material to be located at the top or bottom but to ensure deformation is symmetrical, the patterns will need to alternate between the top and bottom. In addition, the second type of aggregate pattern we demonstrate is obtained through varying material distribution along the thickness direction. This is where all the soft material must occupy all voxels in each layer before moving onto the next layer along the thickness, hence the front and back of the undeformed petals (see the first row in Fig.~\ref{fig:deformed_shape_d3.pdf}(b)) show different colours. Lastly, we show that it is also possible to obtain a homogenised structure by allowing random distribution along both the width and thickness direction. As shown in Fig.~\ref{fig:deformed_shape_d3.pdf}(c), two types of random pattern are considered.

Furthermore, the comparison between the FEM simulations and analytical solutions (see Fig.~\ref{fig:deformed_shape_d3.pdf}(d)) suggests that the designation of material to each voxel has no impact on the overall curvature. This presents additional proof of the adaptability provided by the suggested design approach. Another advantage of employing this design methodology is the ability to fabricate structures using various manufacturing techniques, such as laser cutting, fused filament fabrication or even injection moulding, thereby increasing the practicality of FGC-based shape-morphing structures for industrial applications.

\subsection{Shape morphing structures with different curvature distributions}

\begin{figure}[!h]
   \centering
   \includegraphics[width=1.0\textwidth]{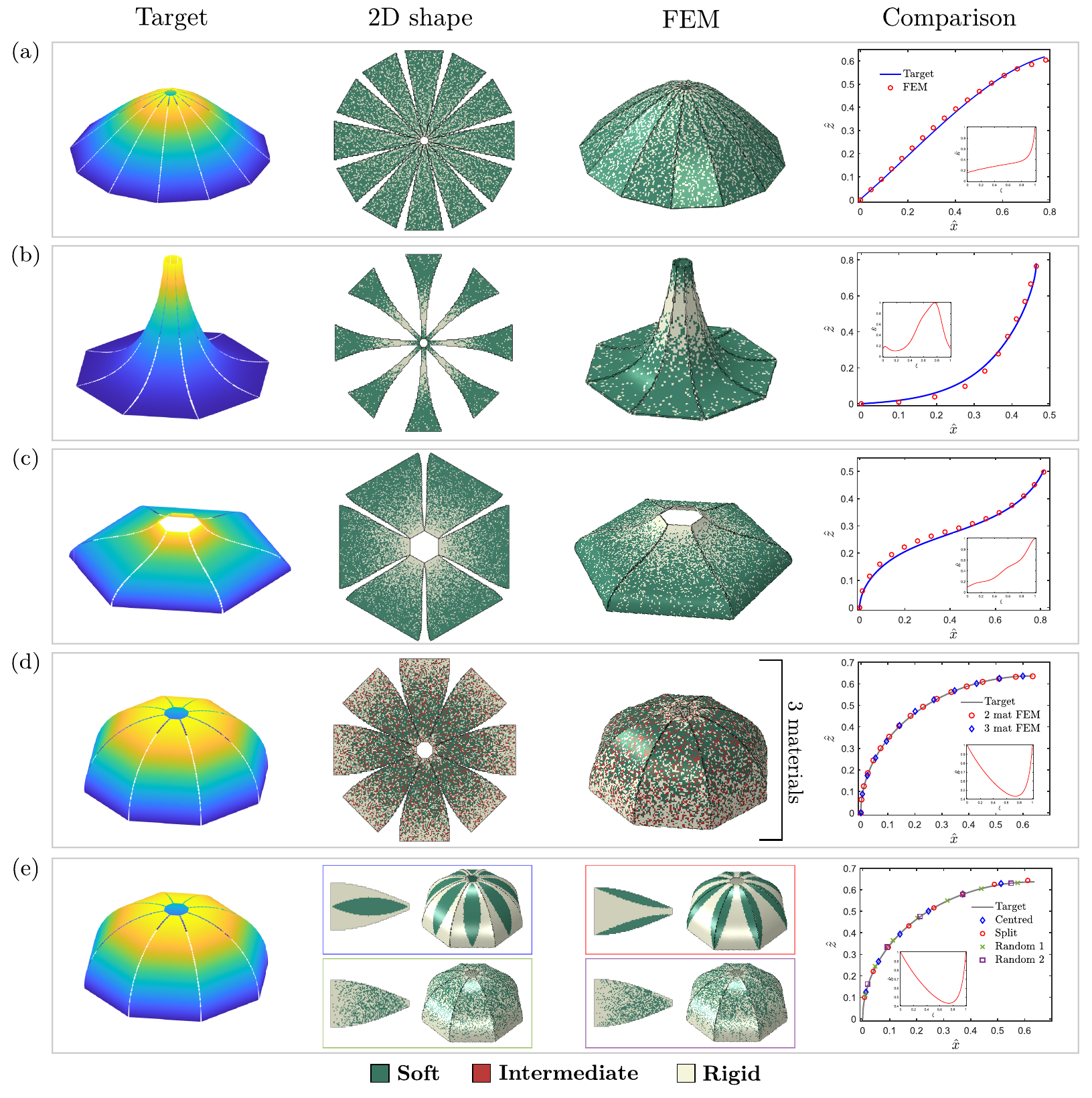}
   \caption{\label{fig:new shapes}Additional examples of shape-morphing structures (a-c) with different distribution of apparent Gaussian curvatures (AGC) or (d) fabricated using three phases of materials. (Left to right) Illustration of the target shape, the 2D flat pattern, and the morphed 3D shapes obtained from FEM simulation, as well as the comparison between the target shape and deformed profile together with a subfigure indicating the modulus grading across each FGC strip. 
   (a) A nose cone structure with positive AGC ($N = 12$, $R = 0.05 $). (b) A trumpet structure with negetive AGC ($N = 8$, $R = 0.05 $). (c) A roof structure with varying AGC ($N = 6$, $R = 0.2 $). (d) A hemisphere formed using three-phase materials ( 
   $N = 8$, $R = 0.1 $) with volume fraction of intermediate phase set as 0.2.}
\end{figure}

In addition to the aforementioned hemispherical morphing structures, the desired target shapes with different distributions of apparent Gaussian curvatures (AGC) can also be generated based on FGCs according to the inverse design framework presented in this work. As shown in the first column in Fig.~\ref{fig:new shapes}(a)-(c), we choose three 3D structures with positive, negative and varying AGC as the target. By giving a particular number of FGC strips $N$ and the inner radius $R$, normalised by the length, and following the similar inverse design procedure as described in Section \ref{section:FEM validation}, we can determine the corresponding 2D cut pattern, as well as the distribution of Young's modulus, as shown in the second column and the insert of the fourth column in Fig.~\ref{fig:new shapes}(a)-(c), respectively. By employing FEM simulation, the morphed shapes of those three designed structures can be obtained (see the third column in Fig.~\ref{fig:new shapes}(a)-(c)). Comparisons between the deformed profiles obtained from FEM simulations and the target shapes are presented in the last column of Fig.~\ref{fig:new shapes}, again showing great agreement.

\label{section:additional examples}

All the examples of FGC-based morphing structures presented before rely on the combination of two phases (soft and rigid) of material. Nevertheless, the capability of implementing multiple phases is also demonstrated in Fig.~\ref{fig:new shapes}(d), whereby an intermediary phase is utilised to bind the soft and rigid phases. Localised stiffening is attained by reinforcing the intermediate phase with the rigid phase, similarly, softening is proceeded with the addition of the soft phase instead. Specifically, we utilised the hemispherical structure (as presented in Section \ref{section:3DTessellation}) but with three materials and compared the profile from the FGC-based morphed structure made of two materials, both of them agree well with the target shape. By employing an intermediate phase, the requirement of compatibility (e.g. interface cohesion and low modulus difference) between soft and rigid phases can be significantly relaxed, as the intermediate phase can be used as a bridging phase for soft and rigid phases. This multi-material approach encourages the use of several distinct materials that are otherwise incompatible with the bi-phase approach.  As a result, it allows the use of multiple materials (three or more materials) in the manufacturing process. Nonetheless, choosing the appropriate volume fraction for the intermediate phase requires careful consideration of various factors, such as mechanical properties, manufacturability, and the compatibility of different materials. The detailed analysis will remain for a further study. Broadly, this exhibits the versatility offered by the proposed modulus grading-based inverse design strategy for morphing structures.

\section{Load-bearing capacities of FGC-based morphed structures}
\label{section: load bearing FGCs}

Considering the structures made of FGCs showing light weight and high stiffness/strength simultaneously \cite{isaac2021review}, many engineering applications would benefit from the use of FGC-based morphing structures such as their use in space structures, as well as building infrastructures and constructions \cite{zhang2022shape}. Therefore, it is essential to evaluate their load-bearing capacities. In this section, we shall take the semi-ellipsoids with different aspect ratios as the candidate to examine both the rigidity at the linear elastic deformation stage and the energy absorption capacity at the non-linear deformation stage.

\subsection{Generation of hemi-ellipsoids with different aspect ratios}

\begin{figure}[!h] 
   \centering
   \includegraphics[width=1\textwidth]{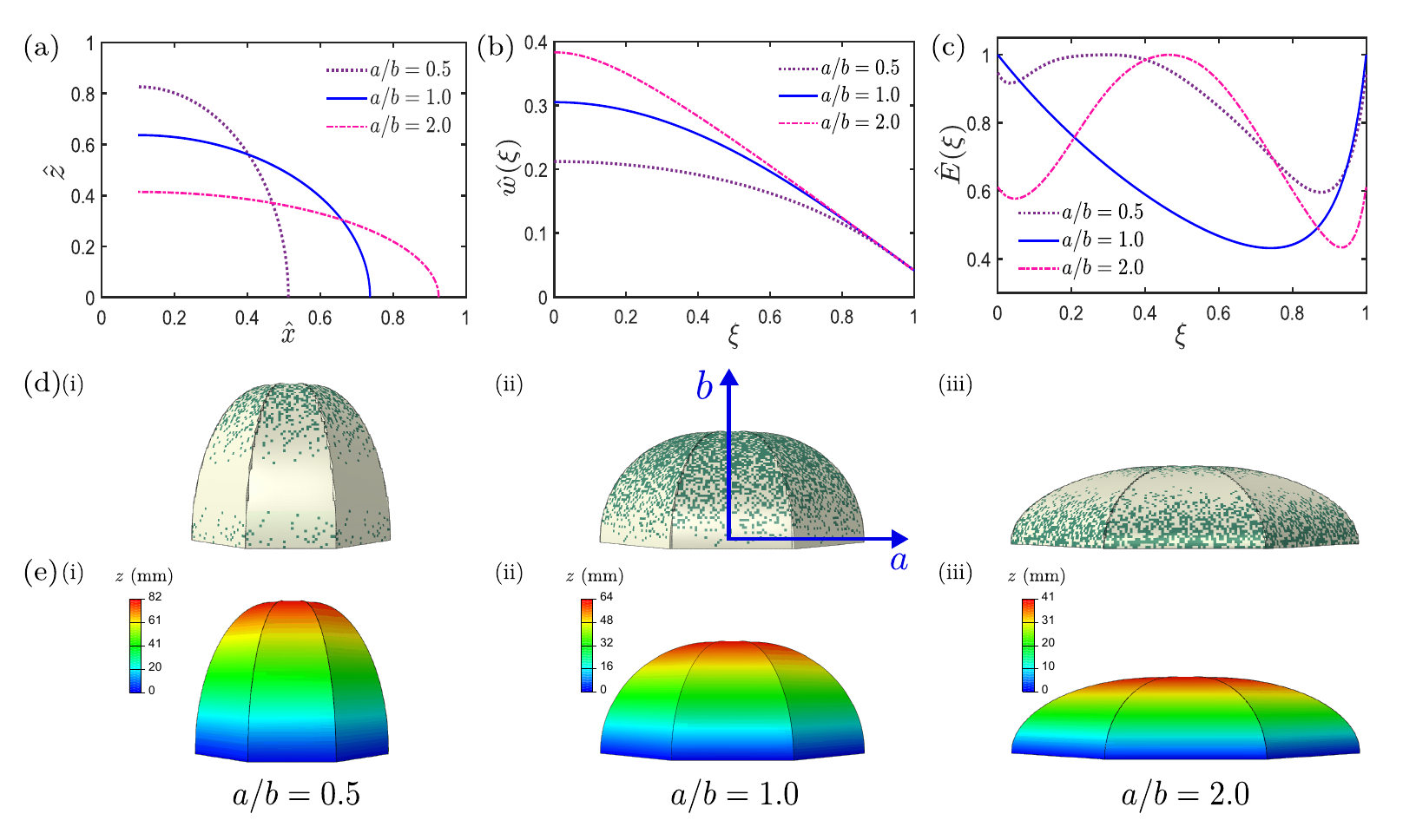}
   \caption{\label{fig:parametric_study_undeformed}Hemi-elliptical structures generated by the inverse design framework with varying aspect ratios ($a/b =$ 0.5, 1.0 and 2.0). (a) The corss-sectional profiles of hemi-ellipsoids with differing aspect ratios; (b) The tapered width profile of each strip depending on the desired 3D shape; (c) The modulus distribution required to reach specific local bending stiffness along the length. (d) Morphed hemi-elliptical structures with different $a/b$ from voxel-based FEM simulations: (i) $a/b = 0.5$, (ii) $a/b = 1.0$, (iii) $a/b = 2.0$. (e) Displacement contour of morphed hemi-elliptical structures obtained from shell-based FEM simulations.}
\end{figure}

In order to quantify the load-bearing capacities of FGC-based morphed structures, we start by considering structures with simple geometry. One potential example is the hemi-ellipsoidal structures, which shape can be described by the elliptic curve, $x^2/a^2 + y^2/b^2 = 1$, where $a$ and $b$ are the major and minor semiaxes, respectively. The geometry of hemi-ellipsoids can be controlled by choosing the semiaxes ratio, $a/b$, in which, the hemispherical morphing structure (see Fig.~\ref{fig:parametric_study_undeformed}) can be considered as a special case, such that $a/b = 1$. Therefore, a large design space of hemi-ellipsoidal structures can be obtained by varying the value of $a/b$.

The arc length of the hemi-ellipsoidal shape can be calculated as \cite{anakhaev2019analytical}

\begin{equation}
    s = a \int\limits^{\varphi}_0\sqrt{1-\left (1-\frac{b^2}{a^2}\right )\sin^2\varphi}d\varphi,
    \label{equ: arc length of ellipse}
\end{equation}
where $\varphi$ is the amplitude function, which corresponds to the ``linear” angle of the parametric equation of the ellipse, i.e. $x = a\sin{\varphi}$ and $y = a\cos{\varphi}$, and can be determined as

\begin{equation}
    \varphi = \arcsin\left[ \frac{b \tan\alpha}{\sqrt{a^2+b^2 \tan^2\alpha}} \right],
    \label{equ: amplitude function}
\end{equation}
where $\alpha$ is the angle between the radius vector to the point under consideration and the horizontal ($x$) axis. Accordingly, the function $\theta(s)$ can be obtained through combining Eqs.~(\ref{equ: arc length of ellipse}) and (\ref{equ: amplitude function}), following $\theta(s) = \arctan\left({dx}/{dy} \right).$

Upon integrating the function $\theta(s)$ with given values of aspect ratio $a/b$, the profiles of hemi-ellipsoids ($\hat{z}$ versus $\hat{x}$) can be obtained, which serve as the target shape. As shown in Fig.~\ref{fig:parametric_study_undeformed}(a), for demonstration, we presented three hemi-ellipsoids with aspect ratios $a/b = 0.5$, 1.0 and 2.0, which were obtained through a choice of $a$ and $b$. Note that here, for given ratio of $a/b$, different values of $a$ and $b$ will give structure with different dimensions. To ensure the morphed structures are comparable, we fix the length of each strip, $L$, as a constant. To implement the inverse design, we take the function $\theta(s)$ as the input information and follow the  procedure presented in Section \ref{section:theoreticalformulation}, the corresponding width distribution of the 2D cut pattern and the modulus distribution for three hemi-ellipsoids can be obtained as shown in Figs.~\ref{fig:parametric_study_undeformed}(b) and (c), respectively.

From a geometric point of view, we can summarise that reducing the aspect ratio yields a smaller strip width (see Fig.~\ref{fig:parametric_study_undeformed}(b)), compensated through an increased height of the morphed structure, as shown in Fig.~\ref{fig:parametric_study_undeformed}(a). The opposite also holds; an increase in strip width results in a shorter morphed structure. Furthermore, the difference in the longitudinal modulus profiles as a result of differing $a/b$ is presented in Fig.~\ref{fig:parametric_study_undeformed}(c). 

To validate the accuracy of the inverse design results, we employed the FEM simulation (following the same procedure described in Section \ref{section:FEM validation}) to implement the morphing process, and the corresponding morphed 3D structures are shown in Fig.~\ref{fig:parametric_study_undeformed}(d) and (e). Note that here, for the results presented in Fig.~\ref{fig:parametric_study_undeformed}(d), we used the voxel-based FEM model, where the green and white elements correspond to soft and rigid materials, respectively. In Fig.~\ref{fig:parametric_study_undeformed}(e), we show the displacement contours for the three morphed hemi-ellipsoids obtained from the shell-based FEM model. It is clearly seen that the morphed shapes obtained from the two different FEM models match each other very well. It is also worth noting that except for those three cases, we can easily implement the inverse design for more target 3D hemi-ellipsoidal shapes with different aspect ratios, and the resulting structures can be used to perform an indentation test for evaluating the load-bearing capacity.

\subsection{The indentation test for the morphed structures}

\begin{figure}[H] 
   \centering
   \includegraphics[width=1\textwidth]{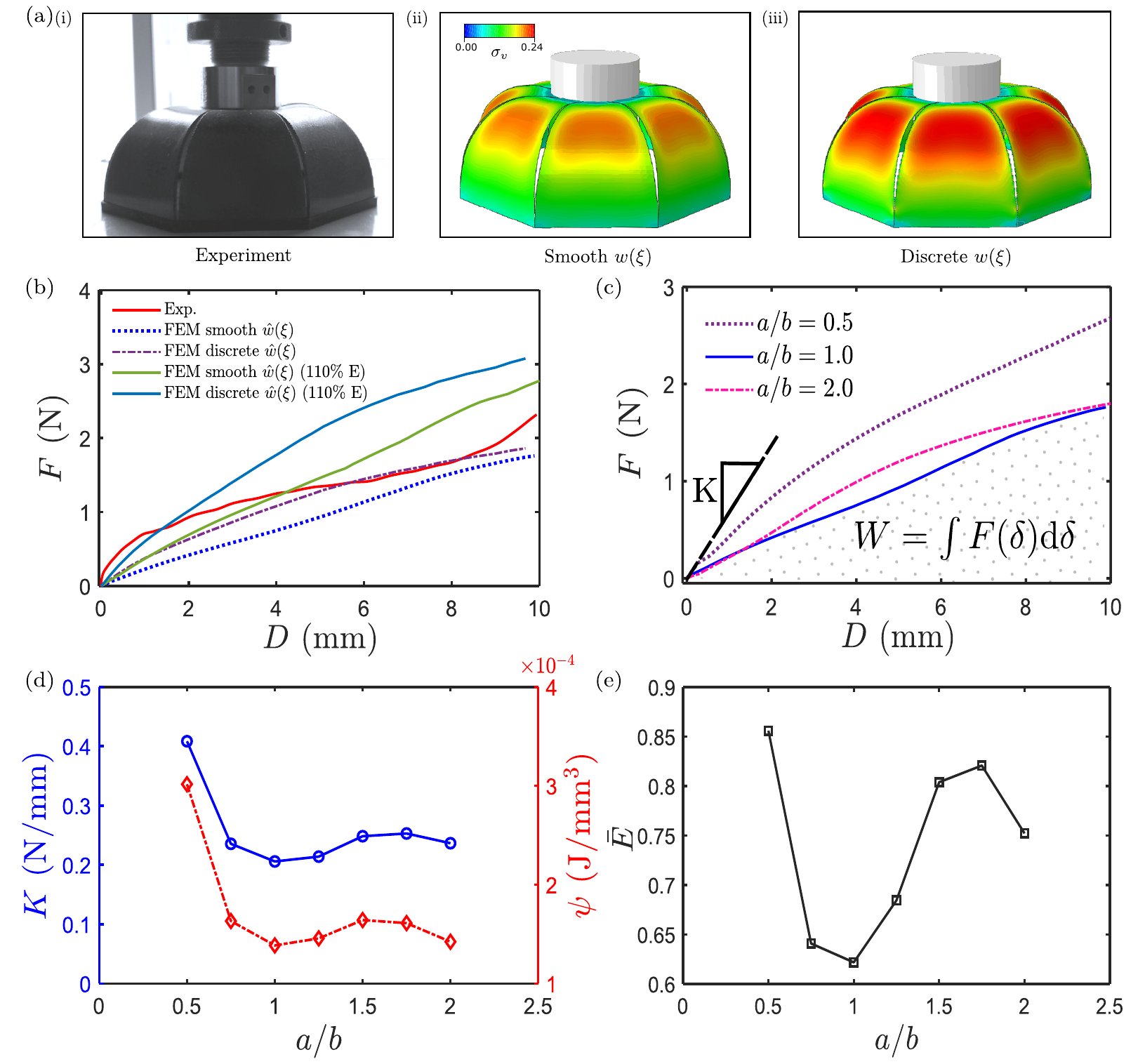}
   \caption{\label{fig:parametric_study_deformed.png} Characterisation of compressive performances (load-bearing capacities) of FGC-based morphing structures.   
   (a) Representation of the indentation of morphed hemisphere based on FGCs in both experiment and FEM simulations: (i) Experimental setup of the indentation test using a flat-tip indenter; (ii) Von Mises stress contour obtained by performing indentation testing using shell-based FEM model smooth width profile; (iii) Stress contour obtained from shell-based FEM simulation with  discrete width profile. (b) {Comparison of Force-Displacement ($F-D$) relations obtained from experiment and two different types of FEM simulations. It further shows the sensitivity of FEM results with a 10\% increase in the Young's Modulus.} (c) Force-Displacement relations extracted from simulation for morphed hemi-ellipsoids with aspect ratios $a/b=$ 0.5, 1.0~\&~2.0, in which the rigidity, $K$, and the specific energy absorption, $\psi$, can be measured accordingly. (d) Comparison of $K$, and $\psi$ for hemi-ellipsoids with a range of aspect ratios extracted from FEM simulation. (e) The relationship between the average modulus $\bar{E}$ and the aspect ratio $a/b$. }
\end{figure}

The indentation behaviours of FGC-based morphing structures under compressive load were investigated in both experiments and FEM simulations. First of all, an indentation test for experimentally assessing the load-bearing capacity was conducted on the hemispherical morphed structure (i.e., $a/b = 1$). Specifically, the experimental setup is shown in Fig.~\ref{fig:parametric_study_deformed.png}(a)-(i): the top central hub of the morphed hemisphere is compressed by a rigid indenter made of stainless steel. The hemispherical shape was deformed using an INSTRON-5967 machine with a flat tip measuring 0.339 in dimensionless diameter, normalised by the length of the shape (147.324 mm), and a stiffness of 190 GPa, significantly stiffer than the printed FGC. The dimensionless deformation depth was 0.157, normalised by the height of the shape (63.662 mm). The compressive load is applied at a strain rate of $0.1~s^{-1}$ using a 100 N load cell. The resulting reaction force, $F$, and displacement, $D$, data were then recorded, and the force-displacement $F-D$ curve is plotted in Fig.~\ref{fig:parametric_study_deformed.png}(b) by a red solid line. The displacement $D$ is measured from recording the crosshead movement of the INSTRON machine. 

In addition, the obtained experimental result was validated by the FEM simulations. Similar to the experimental setup, we performed the indentation test for the morphed hemisphere via simulation. Considering a deep indentation will induce nonlinear deformation and further break the symmetry of the structure, the entire hemispherical structure with eight FGC strips together with the central hub is modelled (rather than model a single strip and employing the symmetric condition, which has been done for modelling the morphing process), as shown in Fig.~\ref{fig:parametric_study_deformed.png}(a)-(ii)$\sim$(iii). Utilising the full FEM model with discretised-solid elements would require a substantial computational cost. However, a significantly lower number of elements is required for the shell-based geometry (the meshing required for the shell-based FEM model is shown in Fig.~\ref{fig:mesh comparison}). Consequently, the alternative shell-based model was utilised.

The performance of the morphed structure under indentation was firstly assessed in FEM simulation by assuming a smooth width profile of FGC strip with shell-based geometry (see Fig.~\ref{fig:parametric_study_deformed.png}(a)-(ii) and Fig.~\ref{fig:mesh comparison}). The corresponding $F-D$ curve is shown in Fig.~\ref{fig:parametric_study_deformed.png}(b) by a blue dotted line. The dissimilarity between the numerical and experimental data is clear and might be attributed to various reasons. The first is the use of linear elastic material in FEM simulation, which may not be able to fully capture the complex constitutive behaviour of the real printed FGCs, for instance, there might be viscoelasticity or hyperelasticity involved. Additionally, the interference between edges of adjacent strips in the printed sample -- the excess width offered by rounding the width values during discretisation causes overlap with adjacent strips when attempting to morph the structure -- is inevitable in our FEM model with a smooth width profile. Consequently, it requires the use of discrete tapering obtained as a result of voxelated geometry. Modelling such geometrical interference within ABAQUS Standard may lead to difficulties in numerical convergence. Alternatively, we employ the general contact implemented in tandem with the use of ABAQUS Explicit solver to deal with the FEM model with discrete width profile (Fig.~\ref{fig:parametric_study_deformed.png}(a)-(iii) and Fig.~\ref{fig:mesh comparison}). The corresponding simulation result, in the form of $F-D$ curve, is also plotted in Fig.~\ref{fig:parametric_study_deformed.png}(b) shown by a purple dash-dotted line. A stiffer response can be noted, in comparison to the FEM model with the smooth tapering profile. However, in order to capture the details of the discrete width profile, a finer mesh with much more (around 8 times) elements is required. Considering the trade-off between accuracy and computational efficiency, we shall use the FEM model with a smooth width profile for further investigations.
Additional simulations also show that a 10\% increase in Young's modulus has a noticeable impact on the stiffness of the morphed structure (see the blue and cyan solid lines in Fig.~\ref{fig:parametric_study_deformed.png}(b)). This observation may suggest that the discrepancies observed between the FEM predictions and the experimental results could be attributed to the uncertainty associated with the initial material characterisation tests.\\

\subsection{Evaluating the compressive performance of FGC-based morphing structures}

Ensuing the validation using $a/b = 1.0$, FEM simulations for hemi-elliptical structures with different $a/b$ have also been performed. In Fig.~\ref{fig:parametric_study_deformed.png}(c), we plotted the $F-D$ curves of three cases with $a/b$ = 0.5, 1.0 and 2.0 as a representative. Consequently, the ability to resist the indentation (i.e. the compressive deformation) is quantified through a measurement of the structure's rigidity, $K$, from the $F-D$ curve, calculated within the initial linear elastic region ($D \lesssim t$, suggested by \cite{lazarus2012geometry,zhang2022shape}). In addition, the work has been done by the external load, $W$, during the full loading regime, which is equal to the energy absorption, calculated by accounting for the area underneath the $F-D$ curve through integration. The volume of solid of the morphed structure, $V$, for each geometry was also calculated by integrating the width profile obtained from the analytical solution and multiplying by the thickness of $t$. The quantity of $W$ was then divided by the volume $V$ of the corresponding structure, giving a value of specific energy absorption (per unit volume), $\psi$. Nonetheless, these parameters served as a means to measure the relative load-bearing capability of each structure, commonly used as a performance indicator in the applications for crash energy absorption \cite{lu2015functionally, isaac2021review}. 

Following the same procedure, more FEM simulations of the indentation test on hemi-elliptical structures with the range of aspect ratios ($0.5 \leq a/b \leq 2.0$) were conducted, aside from the initial three cases. The performance indicators in the form of rigidity $K$ and specific energy absorption $\psi$ against the aspect ratio $a/b$ are presented in Fig.~\ref{fig:parametric_study_deformed.png}(d) by blue and red symbols, respectively. In general, the performance indicators display a unique non-monotonic relationship between stiffness $K$ and aspect ratio $a/b$, also with $\psi$ and $a/b$. In particular, both $K$ and $\psi$ reach much  larger when $a/b$ getting smaller comparing to the cases with $a/b \geq 1$. A similar trend between $K$ and $a/b$ has been observed by Zhang et al. \cite{zhang2022shape}, where the load-bearing capacity of porous morphing structures was evaluated.

One simple potential understanding of these non-monotonic relations is that the load-bearing capacities are positively correlated to the average modulus, $\bar{E} = \int^L_0\hat{E}(\xi)  d\xi / L$, across each strip. As shown in Fig.~\ref{fig:parametric_study_deformed.png}(e), the plot of $\bar{E} - a/b$ also shows the non-monotonic trend and with the minima at $a/b = 1$, which is consistent with the basic trends of $K$ and $\psi$. Noted that this consistency is only intuitive and phenomenological, detailed future studies are required to reveal the foundational mechanism. In addition, we can also expect that these non-monotonic relations could also arise from complex interactions between the geometry and microstructure of the topology. However, through the above analysis, we can get the enlightenment that the load-bearing capacities can be improved through (i) the choice of stiffer material for the two constituent materials and (ii) the use of additional structures such as beams as reinforcement. Nonetheless, it should be emphasised that the purpose of the study is to reveal the robustness of the inverse design strategy to create modulus-graded morphing structures. The analysis of the load-bearing capacities presented here is one of the potential applications of FGC-based morphing structures but is not limited to it.

\section{Multifunctionality of the shape-morphing FGCs}
Apart from serving as load-bearing structures, FGC-based shape-morphing structures also exhibit excellent effective (physical) properties, which can be attributed to the selection of suitable materials based on their specific properties for different applications, including but not limited to thermal and electrical conductivities. The freedom to incorporate multiple materials into the design allows for the potential mitigation of a material's limitations by complementing it with another. In this section, we illustrate how, using simulations conducted through COMSOL Multiphysics, the effective thermal and electrical conductivities of the shape-morphing FGCs can be enhanced by carefully selecting constituent materials in comparison to the utilisation of a single material alone. 

We selected two different materials, as presented in Table \ref{tab:Multifunction material properties} in \ref{section:Multifunctionality of FGCs}, which lists the corresponding material properties. Polymide (PI) + 15\% Graphite composite (referred as soft polymer) was preferred due to its superior electrical conductivity, although it has poor thermal conductivity. In contrast, Short carbon fibre-reinforced PLA (rigid polymer) was chosen for its superior thermal conductivity, despite its poor electrical conductivity. Note that we assume perfect compatibility between the two polymer types, which may not be the case in practice. Furthermore, we also assume that the aspect ratio (length-to-thickness) of each strip is sufficiently large to ensure a bending-dominated behaviour.

As mentioned above, to demonstrate the multifunctionality of FGC-based shape-morphing structures, we simulated heat transfer through pure heat conduction as well as generating an electric potential by applying a current across various types of FGCs. Using COMSOL, transient conduction was implemented, whereby a change in temperature is induced across the length of the strip as a result of applied boundary conditions. The rate of heat flow keeps changing until an equilibrium temperature is reached. {The temperature profile over time and spatial locations is governed by the 3D transient heat conduction equation, which can be derived from a combination of the Fourier's law of heat transfer and the thermodynamic principle of conservation of energy, as}

\begin{equation}
    \nabla^2T + \frac{\displaystyle \dot{q}}{\displaystyle k} = \frac{\displaystyle \rho C_p}{\displaystyle k}\frac{\displaystyle \partial T}{\displaystyle \partial {\mathfrak{t}}},
    \label{equ: 3D heat conduction equation}
\end{equation}
where $k$ is the material's thermal conductivity, $\dot{q}$ the rate of heat generation, $\rho$ the material density and $C_p$ the material's specific heat capacity, with $T$ and ${\mathfrak{t}}$ being temperature and time, respectively. This equation is then utilised by COMSOL to obtain the time-dependent temperature field across the shape-morphing FGC structure.

In addition, the governing equation to calculate the electric currents formed using the continuity equation, whereby the electric charge is conserved in the presence of electric currents. The resulting equation takes the following form,

\begin{equation}
    \vec{\nabla} \cdot \vec{J} + \frac{\partial \rho_c}{\partial {\mathfrak{t}}} = 0,
    \label{equ: electric conduction equation}
\end{equation}
where $\vec{J}$ is the current density, ${\mathfrak{t}}$ the time and $\rho_c$ the charge density. Following Ohm's law, when an electric field is applied to a material, it will induce a flow of current across the material, allowing the following equation to be formed, 

\begin{equation}
    \vec{J} = \bar{\sigma}\vec{{\mathcal{E}}_c} .
\end{equation}

This suggests that the current density, $\vec{J}$, is proportional to the electric field, $\vec{{\mathcal{E}}_c}$, with the constant of proportionality being the material's electrical conductivity, $\bar{\sigma}$. These equations allow the relation between current density and electric field to be explored at a given point, thus allowing the localised behaviour to be analysed. However, when a material is exposed to a uniform electric field and the field is aligned parallel to the length of the material, the potential difference, ${\Phi}$, between both ends of the material can be expressed as ${\Phi} = {\mathcal{E}}_c L$, where $L$ is the length of the material. Since all point act along the length the vector notation can be dropped, Ohm's law reduces to $J = \bar{\sigma} {\mathcal{E}}_c$. Additionally, as ${\mathcal{E}}$ is uniform, the value of $J$ applies everywhere. Consequently, the total current flowing through any point must be equal to $I = JA$ and $R = L/(\bar{\sigma} A)$, with $A$ being the cross-sectional area of the material. These results can then be combined to the macroscopic form of Ohm's law,
\begin{equation}
    {\Phi} = IR = IL/\bar{\sigma} A.
    \label{equ: macroscopic Ohm's law}
\end{equation}


\subsection{Multifunctional rectangular FGCs}

\begin{figure}[!ht] 
   \centering
   \includegraphics[width=1\textwidth]{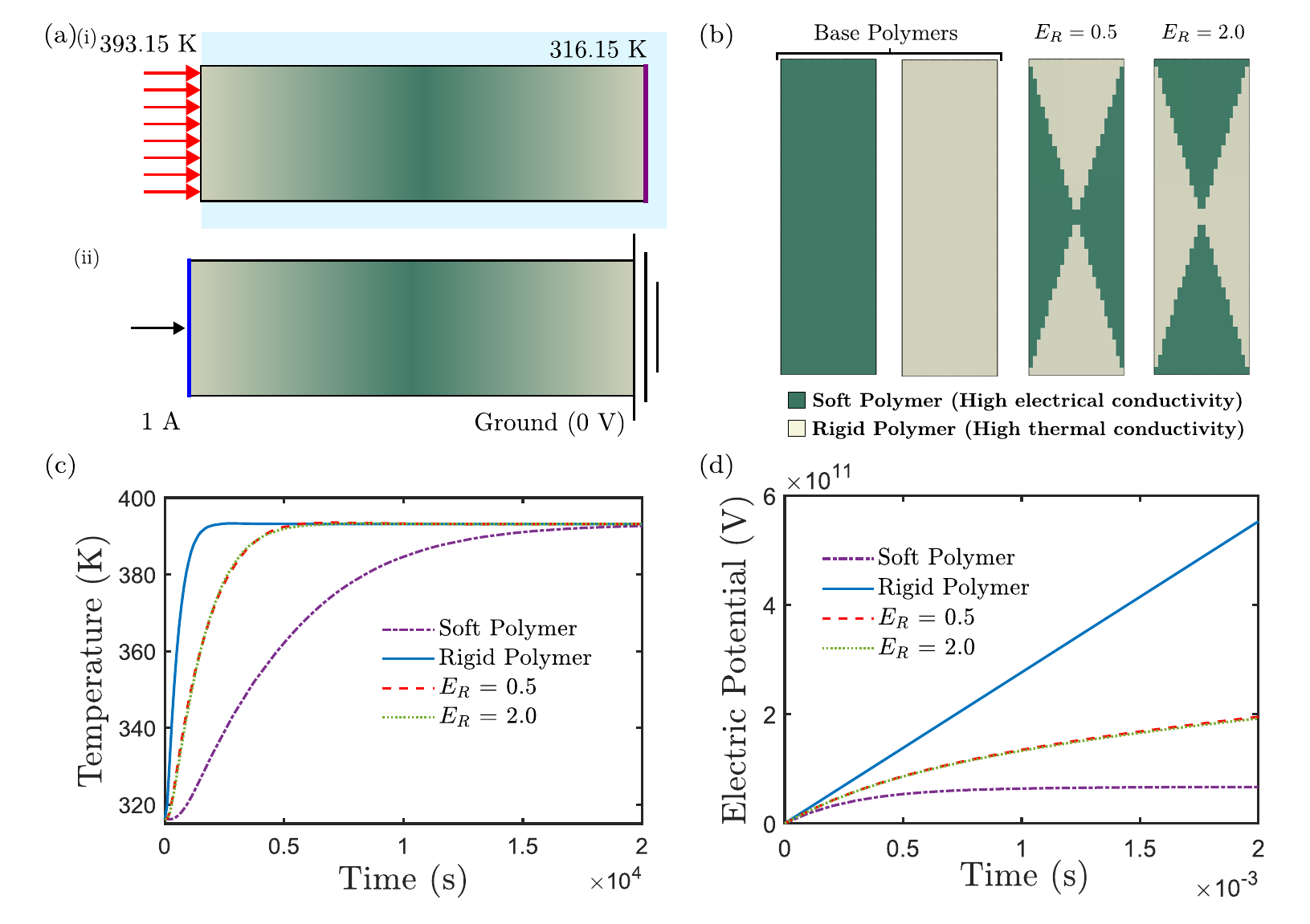}
   \caption{\label{fig:rec_multifunction_d3.pdf} Multifunctionalities of rectangular FGCs. (a) The boundary conditions applied to the rectangular strips for (i) heat transfer and (ii) electric current simulations. (b) Two rectangular FGC strips with different material's distribution together with two strips with single material. FEM results for (c) heat transfer and (d) electric current simulations. Each strip here has a height of 40 mm and width of 12 mm.}
\end{figure}

{To evaluate the multifunctionality of FGCs, we commence by examining rectangular strips with different compositions, as illustrated in Fig.~\ref{fig:rec_multifunction_d3.pdf}, through numerical simulation in COMSOL. Also illustrated in Fig.~\ref{fig:rec_multifunction_d3.pdf}(a) are the boundary conditions, imposed at relevant points depending on the type of simulation performed.} In the heat transfer simulation, a constant source of heat at temperature 393.15 K is applied to the left side of the strip, which is immersed in an environment at a temperature of 316.15 K. A vertical purple line indicates the location where the results were obtained using a surface average. In the electric current simulation, a current of 1 A is supplied across the left edge, while the right side is grounded. The dark blue vertical line represents the location where the electric potential was obtained. {As shown in Fig.~\ref{fig:rec_multifunction_d3.pdf}(b), two FGC strips with aggregate patterns are considered, while the limits of electrical and thermal conductivities were determined by examining instances where the strip was produced using single material, i.e. the base polymers.}

The FEM simulation outcomes for transient heat transfer for four strips are presented in Fig.~\ref{fig:rec_multifunction_d3.pdf}(c). The results exhibit an anticipated trend, i.e., the soft polymer with the lowest conductivity takes the longest time to attain equilibrium, while the rigid polymer is the fastest. Additionally, it is evident that the composite cases ($E_R = 0.5, 2.0$) lie between the two extreme base polymer cases. This is reasonable since there is an approximately 50\%-50\% composition between the two material types. The same trend is noticeable in the electric current simulations {(Fig.~\ref{fig:rec_multifunction_d3.pdf}(d))}, although with one difference - the rigid polymer, which has a lower electrical conductivity than the other base polymer, results in a higher electric potential being generated. However, the total duration for both simulation types was determined by the time taken to achieve the equilibrium state. For example, it took roughly 10,000 seconds for the soft polymer case to reach 393.15 K, assuming negligible heat loss. In the case of electric current simulations, the total runtime was based on the soft polymer, where the magnitude of electric potential plateaued, unlike the rigid polymer, where the magnitude of electric potential would have been much higher with prolonged simulation time. The contour graphs for both of these simulations can be found in \ref{section: rectangular FGC multifunction}, which also shows how the temperature/electric potential varies across the length of each rectangular strip.

\subsection{Multifunctional FGC-based hemispheres}
\begin{figure}[!h] 
   \centering
   \includegraphics[width=1\textwidth]{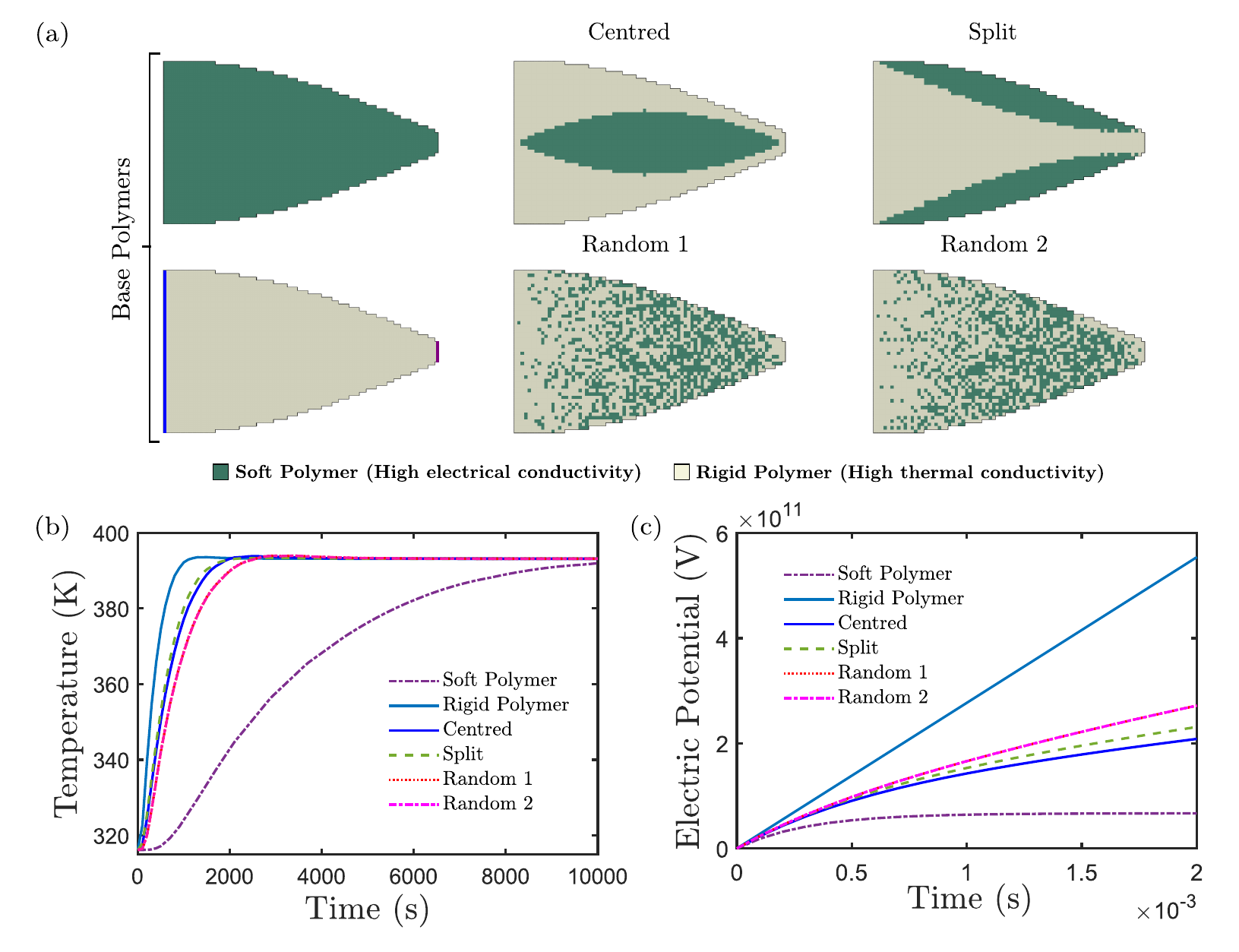}
   \caption{\label{fig:/hemisphere_multifunction_d4.pdf} Multifunctionalities of shape-morphing FGCs. (a) Multiple aggregate patterns used in the multifunctional simulations. The location where electric potential is calculated is marked as \textbf{A}, while the location for temperature measurement is marked as \textbf{B}. (b) The FEM results for heat transfer simulation are presented in this subfigure, covering a time period of 10,000 s. (c) The FEM results for heat transfer simulation are presented in this subfigure, covering a shorter time period of 0.002 s. Each strip here has a length of 40 mm and height of 40$\hat{w}(\xi)$ mm.}
\end{figure}

{We proceed to showcase the multifunctionality of FGCs through a more intricate hemispherical shape consisting of four distinct cases with different material distribution, as shown in Fig.~\ref{fig:/hemisphere_multifunction_d4.pdf}(a), while the two cases with base polymers serve as the upper and lower bounds. A trend similar to the one observed with rectangular FGCs is evident in this case, seen in Fig.~\ref{fig:/hemisphere_multifunction_d4.pdf}(b) and (c), for heat transfer and electric potential generation respectively. The temperature and electric potential values for the FGC structures made by two materials are located between the soft and rigid polymers' set bounds, as expected.} Nonetheless, there are slight variations in the trends for the four FGC cases, which may arise from the distribution of material as well as the location at which the material properties have been measured, labelled ``\textbf{A}" and ``\textbf{B}" or deviations in the volume fractions due to rounding. Nevertheless, the contour plots for each case can be found in \ref{section: hemisphere FGC multifunction}, which also shows the structure's ability to morph as well as its ability to possess superior thermal and electrical properties achieved through strategic material selection.

\subsection{Effective conductivities of shape-morphing FGCs}

\begin{figure}[!ht] 
   \centering
   \includegraphics[width=1\textwidth]{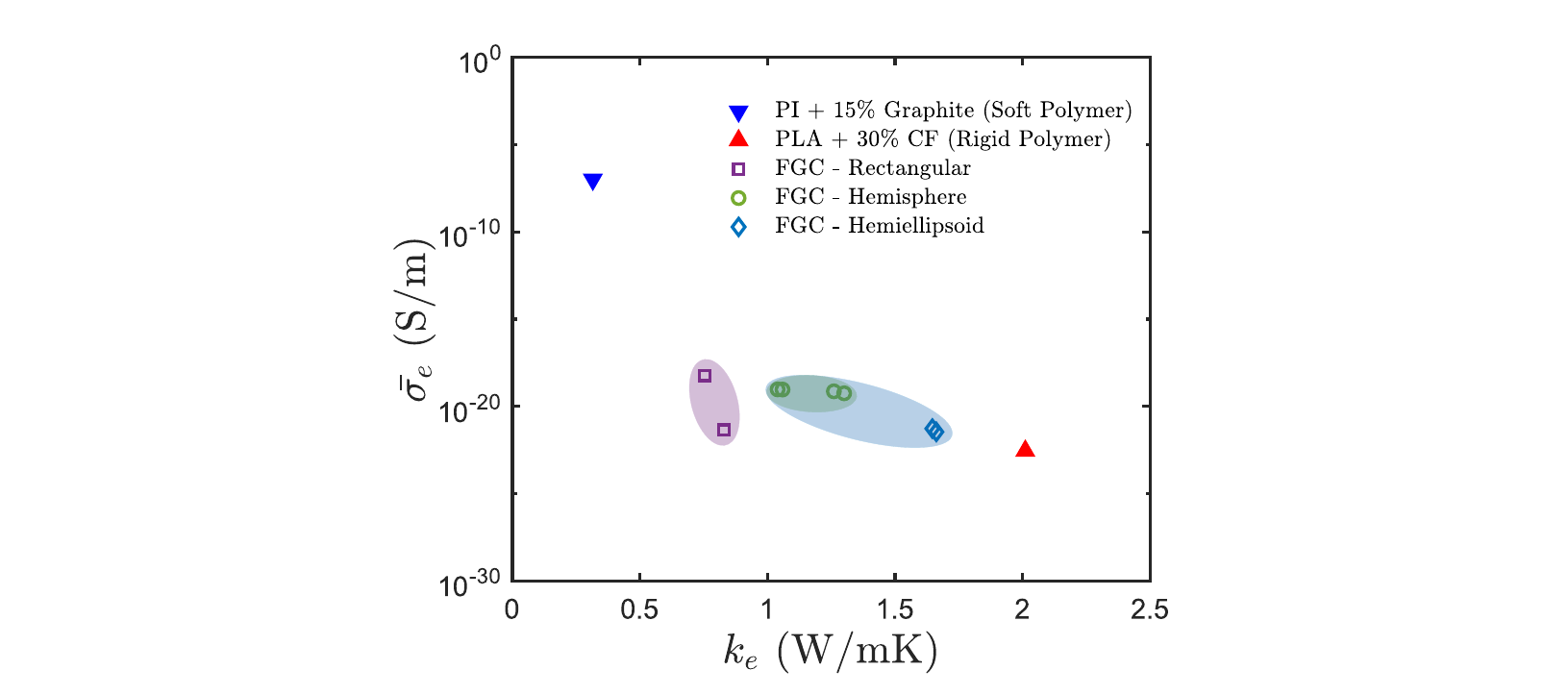}
   \caption{\label{fig:/effective_conductivities_log_log_d4.pdf} {The effective thermal conductivities are plotted against the effective electrical conductivities of shape-morphing FGCs with different geometries and pure soft and rigid polymers.}}
\end{figure}

{To evaluate the overall physcial properties for FGCs, it is crucial to quantify the effects of composition and distribution of different phases of materials. Various homogenization methods, such as the rule of mixtures, Maxwell's effective medium theory \cite{markov2000elementary, maxwell1881treatise}, Mori-Tanaka model \cite{benveniste1987new}, and Hashin-Shtrikman bounds \cite{hashin1963variational}, are widely adopted to obtain homogenised conductivities. Here, to incorporate more physical meaning, we use a numerical approach to estimate the effective thermal and electrical conductivities of FGCs, based on the Fourier's law (i.e. Eq. \ref{equ: 3D heat conduction equation}) and the Ohm's law (Eq. \ref{equ:  macroscopic Ohm's law}), respectively. The details of this approach is provided in \ref{section: Effective conductivities of FGCs}. The results from this approach are plotted in the form of Ashby chart, illustrated in Fig.~\ref{fig:/effective_conductivities_log_log_d4.pdf}. It is clearly shown that, by combining two different materials, each with favourable characteristics, such as one material with high electrical conductivity but low thermal conductivity and opposite for another, one can achieve the combined effective properties for both electrical and thermal conductivities in a single structure made by FGCs. This represents a major benefit of using FGC-based morphing structures since it enables the coexistence of two  distinct advantages of material properties in a single structure.}

\section{Discussion and Conclusion}
This paper presents a novel framework to inversely design shape-morphing composite structures through the introduction of modulus grading and multi-material additive manufacturing. Tapering pattern to fulfil both the tessellation condition and modulus grading for required bending stiffness are predicted through the analytical model based on the theory of graded elastica. Modulus grading was achieved by discretising the resultant geometry, followed by the use of the rule of mixtures to obtain the volume fraction required to get a certain cross-sectional modulus. This also allows the modulus-graded composites (i.e. FGCs) to be fabricated by additive manufacturing (multi-material 3D printing). The design principle is validated using a combination of experiments and numerical analysis with a hemispherical structure being the target shape. Emphasis should be made on the fact that the framework established here is generic for many other geometries, independent of shape and size. As a demonstration, we presented three examples with different curvature distributions to show the generality of our design framework. Except for the two-material additive manufacturing, we also conceptually proved that we can harness more (such as three) materials for manufacturing the FGC-based morphing structures. A further evaluation of the FGC-based hemi-ellipsoidal morphing structures for load-bearing capacities has been performed with varying aspect ratios. The rigidity and specific energy absorption measured from indentation tests were utilised as performance indicators. An investigation was performed through compressive testing of 3D tessellated structure and compared with FEM simulation results. The data obtained revealed a non-monotonic relationship between the rigidity and aspect ratio, with similar results for specific energy absorption. We also demonstrate how the use of two materials in FGCs can effectively overcome the drawbacks of using a singular material with limit of physical properties for multiphysics applications, where through material choice, we can tailor the properties based on requirements.

The inverse design and additive manufacturing framework for the FGC-based shape-morphing structures demonstrated here have many potential applications in the field of engineering, including but not restricted to shape-changing soft robots, deployable space structures and morphable flexible electronics. This preliminary investigation can serve as a guide for the design and manufacture of composite-based shape-morphing structures. {It is important to note that this method is not confined to 3D printing, which may have resolution limitations. Photolithography, as one typical example, can be employed to broaden the scope of this approach to encompass micro-/nano-scale shape-morphing structures. Through careful design and control of the exposure and development processes, it becomes feasible to achieve shape-morphing FGCs at smaller scales.} This study also motivates further studies in other methods of achieving modulus grading, such as fibre orientation as well as the application of actuation. This may be accomplished by embedding stimuli-responsive material within the print or as an addition to the printed structure for shape-morphing to be achieved.

\section{Acknowledgement}
W. Tan acknowledges the financial support from the EPSRC New Investigator Award (grant No. EP/V049259/1). M. Liu acknowledges the support from the Nanyang Technological University via the Presidential Postdoctoral Fellowship. We are grateful to K. Jimmy Hsia and Zhaohe Dai and Dominic Vella for useful discussions.

\section{Data availability}
The data and codes needed to reproduce and evaluate the work of this paper are available in the GitHub repository, once this manuscript is published: (https://github.com/MCM-QMUL/MorhComp.git).

{\footnotesize
\bibliographystyle{unsrt}
\bibliography{bibliography.bib}}

\pagebreak 

\appendix
\setcounter{figure}{0}    
\section{Appendices}

\subsection{Methods of joining two types of polymers}
\label{section:joining polymers}

Many techniques that can be used to achieve compatibility between two different materials for creating compliant materials that can deform without failure. Among them, three typical techniques are presented here as examples.

\begin{figure}[!h] 
   \centering 
   \includegraphics[width=1\textwidth]{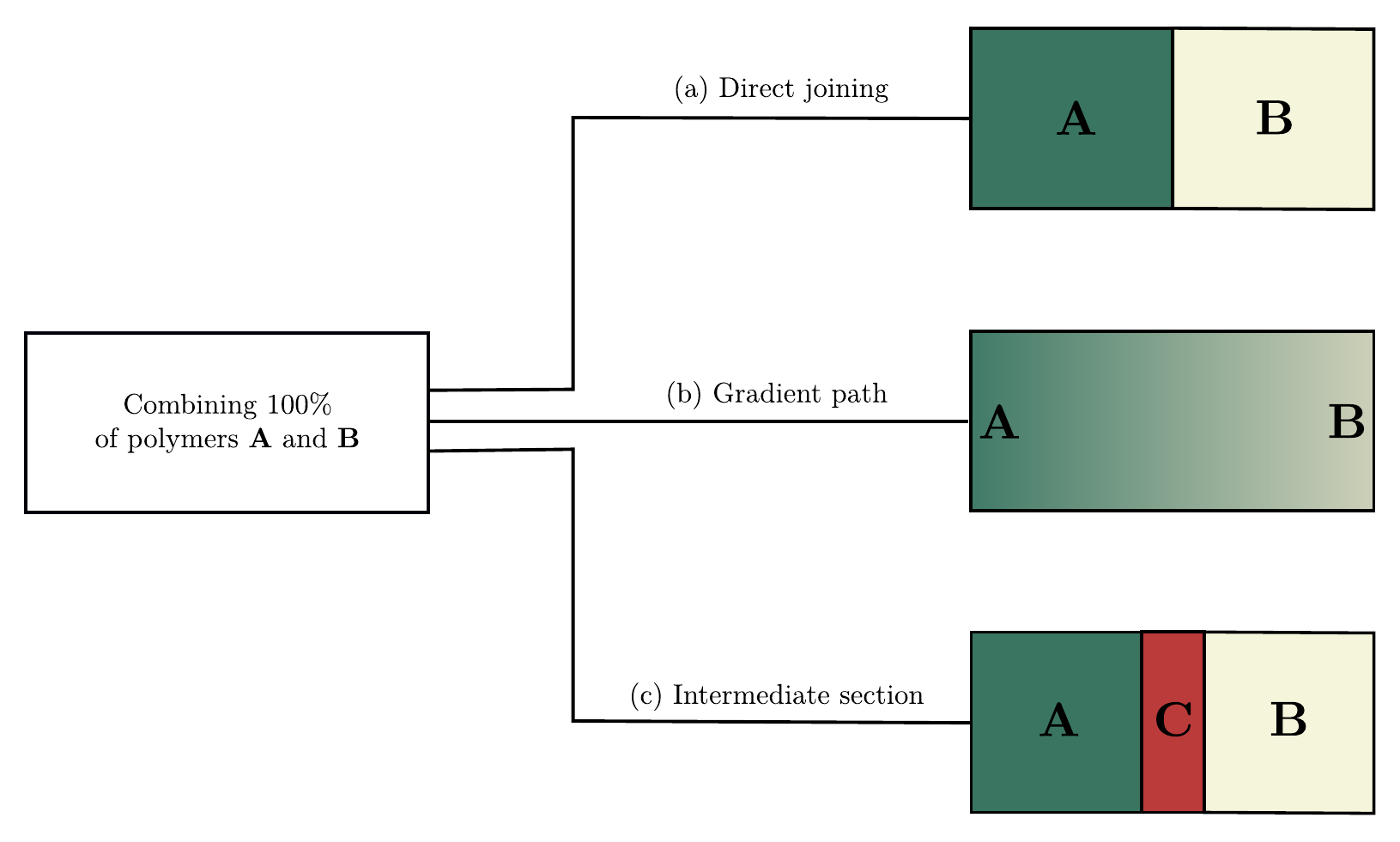}
   \caption{\label{fig:combining two polymers}Various methods of joining two types of polymers together, namely polymers \textbf{A} and \textbf{B}. (a) Direct joining involves physically bonding two different types of polymers together, without the use of additional bonding agents. (b) The gradient path approach describes the process of changing the composition of materials, such that it facilitates bonding between the two materials. The properties change gradually, from 100\% polymer \textbf{A} to 100\% polymer \textbf{B}. (c) Using an intermediate section as a bonding agent between the two materials. This section is made from a material that is compatible with both of the polymers.}
\end{figure}

\subsection{Young's modulus of the candidate polymer materials used in this study}
\label{section:resin modulus}
Coupons were printed using composite resins with varying proportions of ``angoBlack" and ``VeroClear" resins. Tensile testing was conducted using these samples for material characterisation at a rate of 1 mm/min. The stress-strain data obtained was used to calculate Young's modulus and yield strength of each composite resin, with FLX9070 being chosen as the soft material and FLX9095 chosen to represent the rigid polymer. 

\begin{figure}[!h]
    \centering
    \includegraphics[width=1\textwidth]{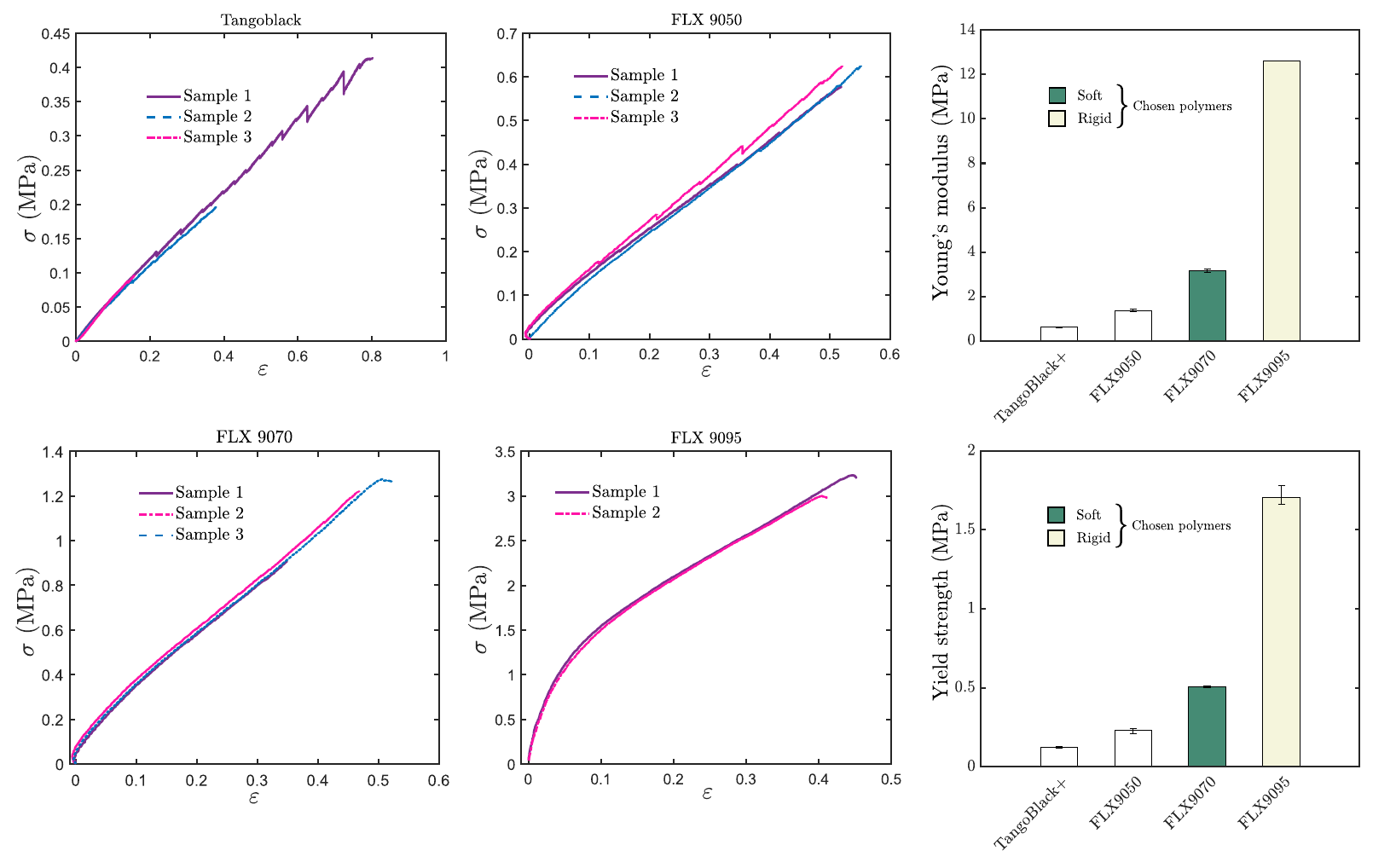}
    \caption{\label{fig:resin modulus}Stress-Strain curves of candidate resins, with Young's modulus and yield strength extracted from tensile test conducted on coupons fabricated using various composite resins presented in the bar chart. Note that the yield strengths of materials are determined by finding the intersection of the stress-strain curves and the 0.2\% offset line parallel to the initial slope.}
\end{figure}

\subsection{Comparison of mesh over smooth and discrete width profiles}
\label{section: mesh comparison}

Here, we show the details of the finite element mesh for different models used in this work. A structured mesh was produced over the discretised-solid model with each voxel being meshed using eight elements. The modulus-graded strip meshed using 221376 C3D8R type elements of size $\eta/2$ and the mesh across the thickness direction being required. The ensemble of graded strips to form the morphing structure would consequently require 1771008 elements. Furthermore, swept mesh was utilised over a smooth-width profile shell model with 16340 S4R elements in total. Lastly, structured mesh over discrete shell geometry with elements of size = $\eta$ with a total of 133840 S4R type elements was obtained by partitioning the geometry. The width and length of each geometry are of magnitude $100 \hat{w}(\xi)$ mm and 100 mm, respectively. It should be noted that the discretised-solid model was utilised to model the shape-morphing by way of obtaining a visual representation of modulus grading over the morphed structure. The shell-based discrete-width profile model was strictly used to model interference (hard contact) between adjacent strips during indentation. Finally, the shell-based smooth-width profile model was used as a computationally-effective alternative to performing both shape-morphing and indentation tests.

\begin{figure}[!h]
   \centering
   \includegraphics[width=1\textwidth]{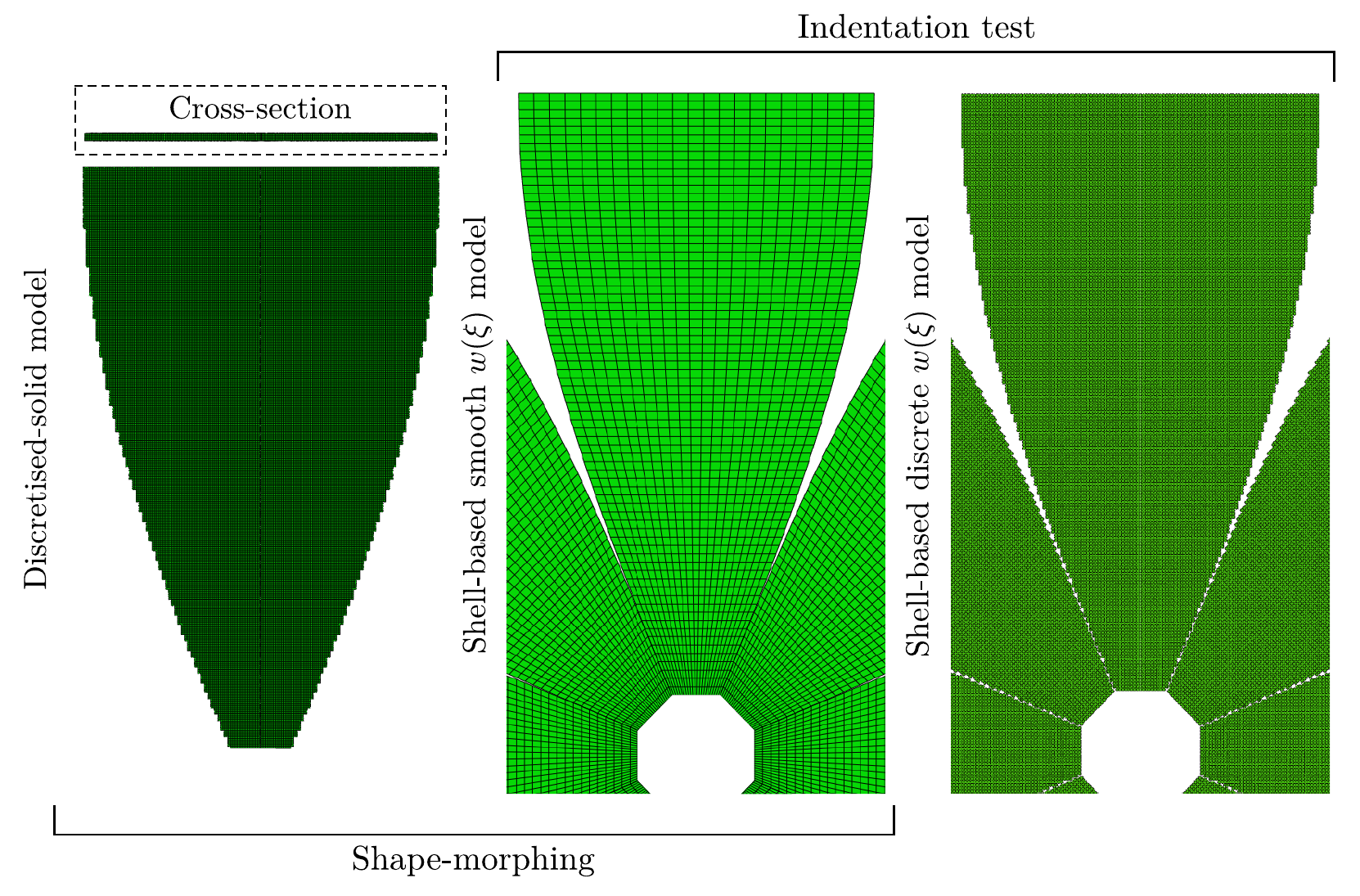}
   \caption{\label{fig:mesh comparison}Comparison of the mesh obtained over hemispherical morphing structure when implementing: (Top to bottom) discretised-solid model, shell-based smooth width profile model, and shell-based discrete width profile model. A single strip of the discretised-solid model, based on an axisymmetric boundary condition, is used to predict the shape-morphing. A full shell-based model with a smooth width profile is used as a computationally-inexpensive method for both the shape-morphing step and indentation test. Note that all eight strips are modelled in the shell-based model. Similarly, the full shell-based model with a discrete width profile is utilised for exploring the interference between adjacent strips during the indentation test.}
\end{figure}


\subsection{Demonstration of multifunctionality exhibited by shape-morphing FGCs}
\label{section:Multifunctionality of FGCs}

In order to demonstrate the multifunctionality of the FGC-based shape-morphing structures via the simulation in COMSOL, we choose two materials with diacritical properties as the candicates for creating FGCs. The corresponding materials' parameters used for simulations are listed in Table \ref{tab:Multifunction material properties}.

\begin{table}[htbp]
  \centering
  \caption{Table of values used to define material properties in simulating heat transfer and electric currents in COMSOL \cite{ansys2023}. \label{tab:Multifunction material properties}}
    \begin{tabular}{lccc}
        \toprule
\multicolumn{1}{c}{Material Property} & \multicolumn{1}{c}{\begin{tabular}[c]{@{}c@{}}Polymide (PI) + \\ 15\%  Graphite \end{tabular}} & \multicolumn{1}{c}{\begin{tabular}[c]{@{}c@{}}Polylactic acid (PLA) + \\ 30 \%  Carbon fibre \end{tabular}} \\
        \midrule
          Young's Modulus (MPa) & 3330 & 10200           \\
          Density (kg/m\textsuperscript{3}) & 1410 & 1270            \\
          Thermal conductivity (W/mK)           & 0.3155 & 2.01         \\
        Electrical conductivity (S/m)           & 1E-7                                                                                                         & 2.86E-23 \\

        \bottomrule
    \end{tabular}
\end{table}

\subsubsection{Multifunctionality of rectangular FGCs}
\label{section: rectangular FGC multifunction}
The following section contains the contour plots obtained when running heat transfer and electric current simulations on COMSOL using the rectangular FGCs.

\begin{figure}[!h] 
   \centering
   \includegraphics[width=1\textwidth]{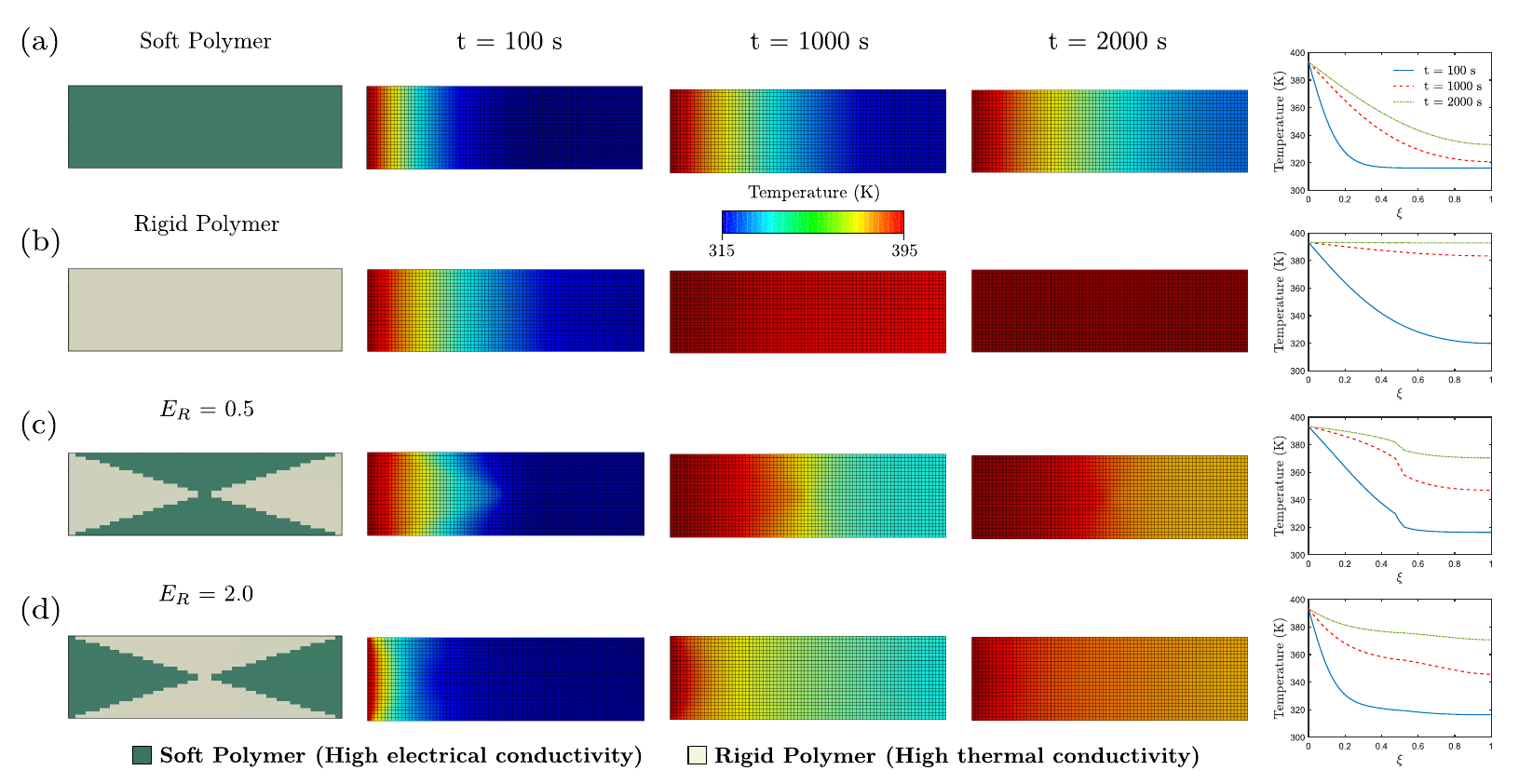}
   \caption{\label{fig:/rec_heat_transfer_images_d3.pdf} Contour plots depicting the heat transfer occurring over time along the length of rectangular structures made of FGCs are shown. The left column displays the composition of each FGC structure, while the middle columns displaying the contour plots at t $= 100, 1000, 2000$ s. The right column shows the heat transfer over the length of the graded strip at various times.}
\end{figure}

\begin{figure}[!h] 
   \centering
   \includegraphics[width=1\textwidth]{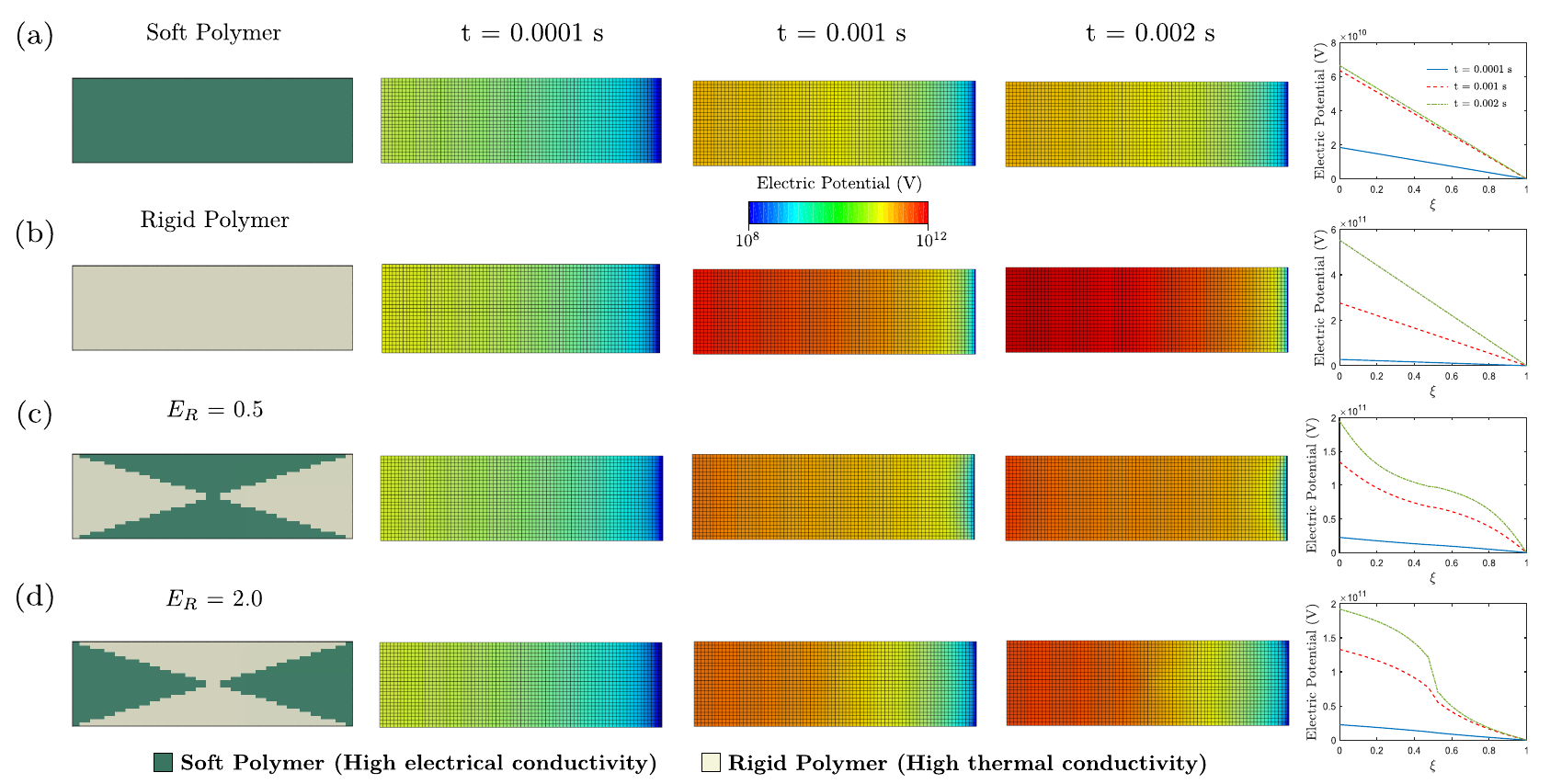}   \caption{\label{fig:/rec_electric_potential_images_d2.pdf} The figures exhibit contour plots portraying the evolution of electric fields over time along the length of rectangular FGC-based structures. In the left column, the composition of each FGC structure is presented, while the middle columns exhibit the contour plots at t $= 0.0001, 0.001, 0.002$ s. The right column illustrates the electric potential over the length of the graded strip at different times.}
\end{figure}

\subsubsection{Multifunctionality of hemispherical FGCs}
\label{section: hemisphere FGC multifunction}

Here, we show the results obtained from COMSOL, where simulations were conducted on hemispherical structures with aggregate patterns, to see its effect on heat transfer and the  electric potential generated.

\begin{figure}[!h] 
   \centering
   \includegraphics[width=1\textwidth]{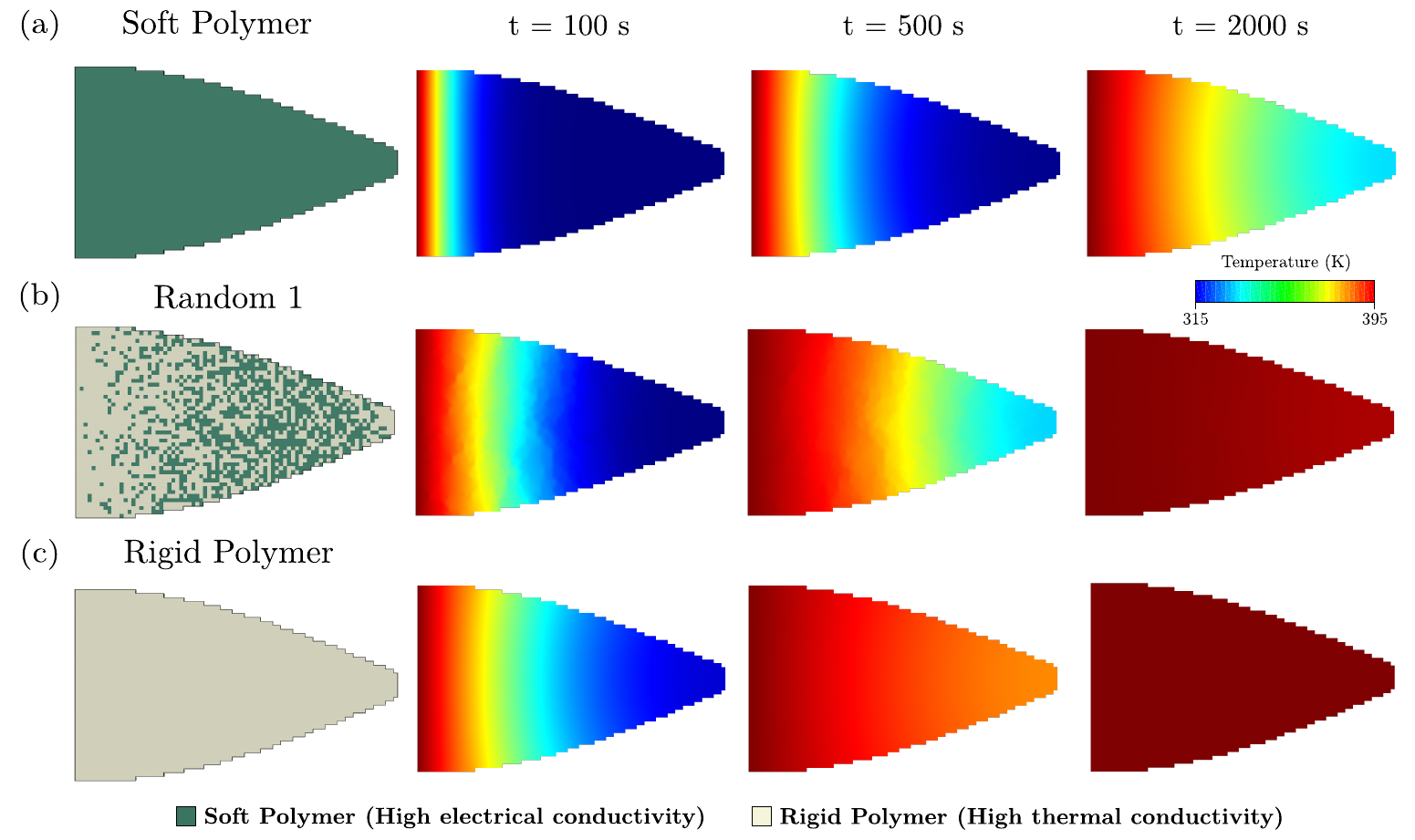}   \caption{\label{fig:/hemisphere_heat_transfer_images_petal.pdf} Contour plots resulting from heat transfer occurring on singular petals. It should be noted that here we use a singular petal as a single material structure when morphed will not achieve the target shape. The first column shows the material composition of each hemispherical strip and the successive columns show the temperature contour at time intervals t $= 100, 1000, 2000$ s.}
\end{figure}

\begin{figure}[!h] 
   \centering
   \includegraphics[width=1\textwidth]{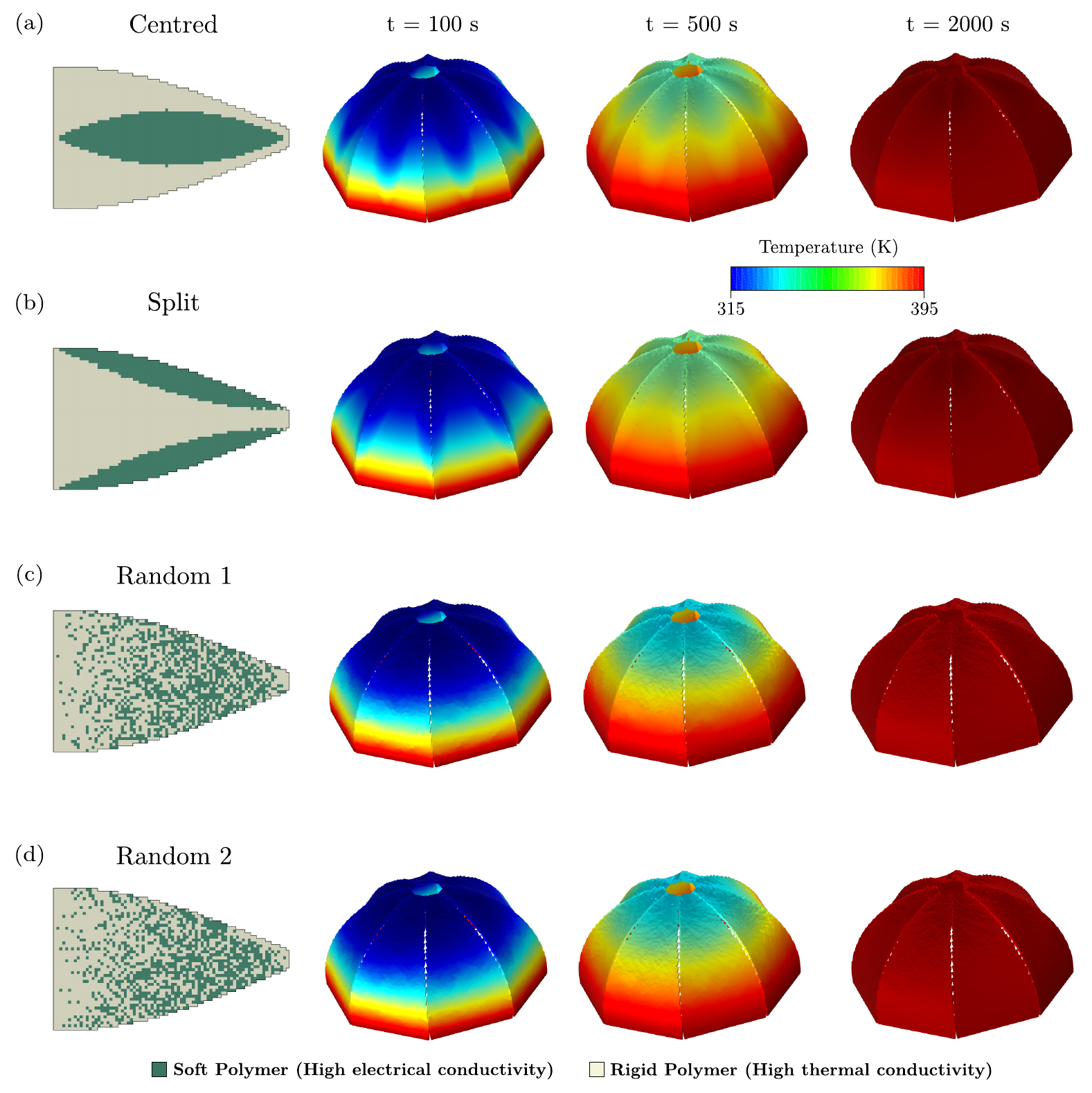}   \caption{\label{fig:/hemisphere_heat_transfer_images_full_hemisphere.pdf}  The contour plots illustrate the heat transfer occurring across deformed hemispherical structures. The first column shows the material composition of each hemispherical strip, whereby a complete hemispherical structure is comprised of eight different strips and the successive columns show the temperature contours at time intervals t $= 100, 1000, 2000$ s.}
\end{figure}

\begin{figure}[!h] 
   \centering
   \includegraphics[width=1\textwidth]{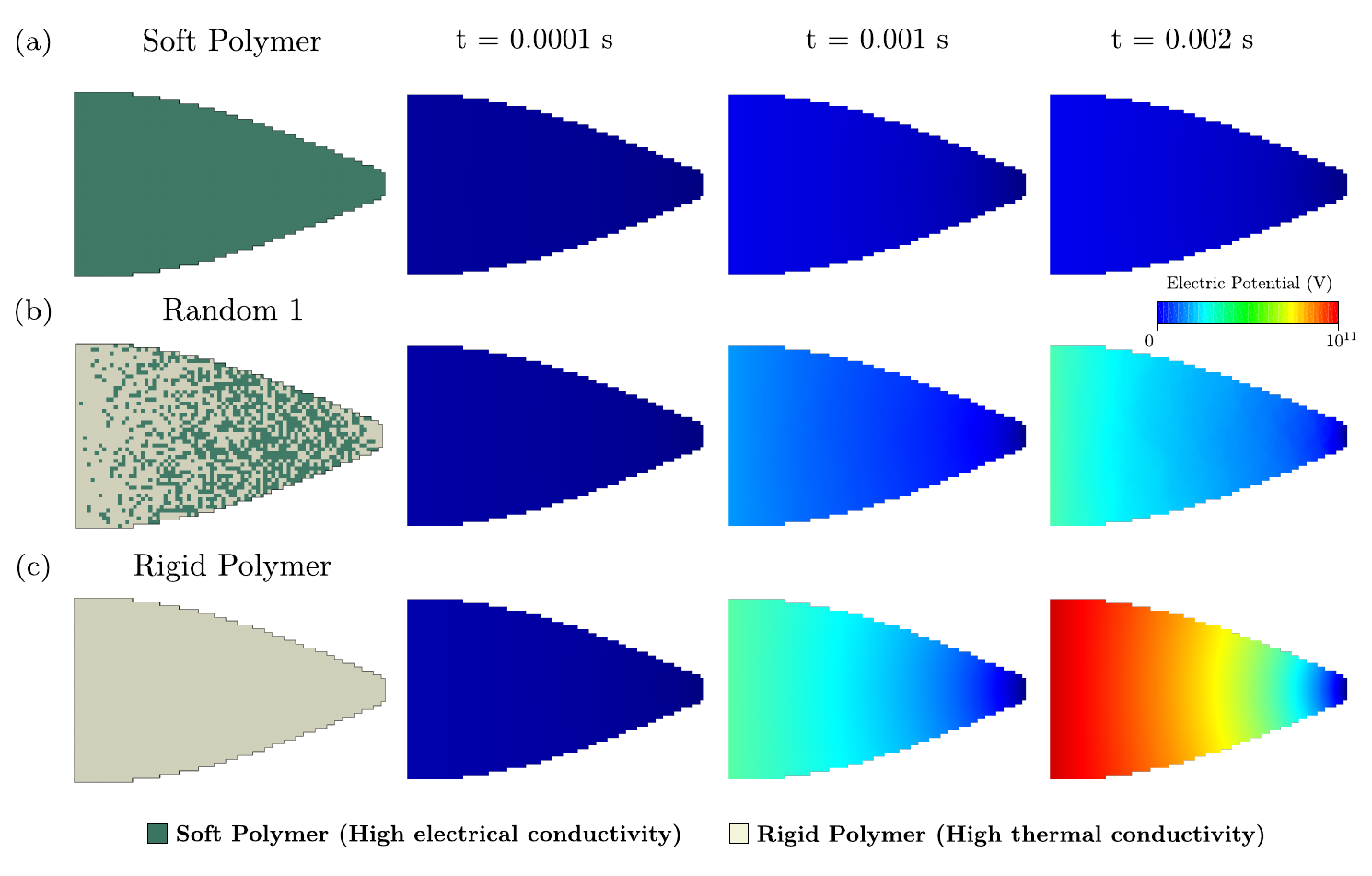}   \caption{\label{fig:/hemisphere_electric_current_images_petal.pdf} These contour plots depict the electric current distribution on a singular petal. The first column illustrates the material composition of each hemispherical strip. The subsequent columns show the contour plots of electric potential at time intervals of t $= 0.0001, 0.001, 0.002$ s.}
\end{figure}

\begin{figure}[!h] 
   \centering
   \includegraphics[width=1\textwidth]{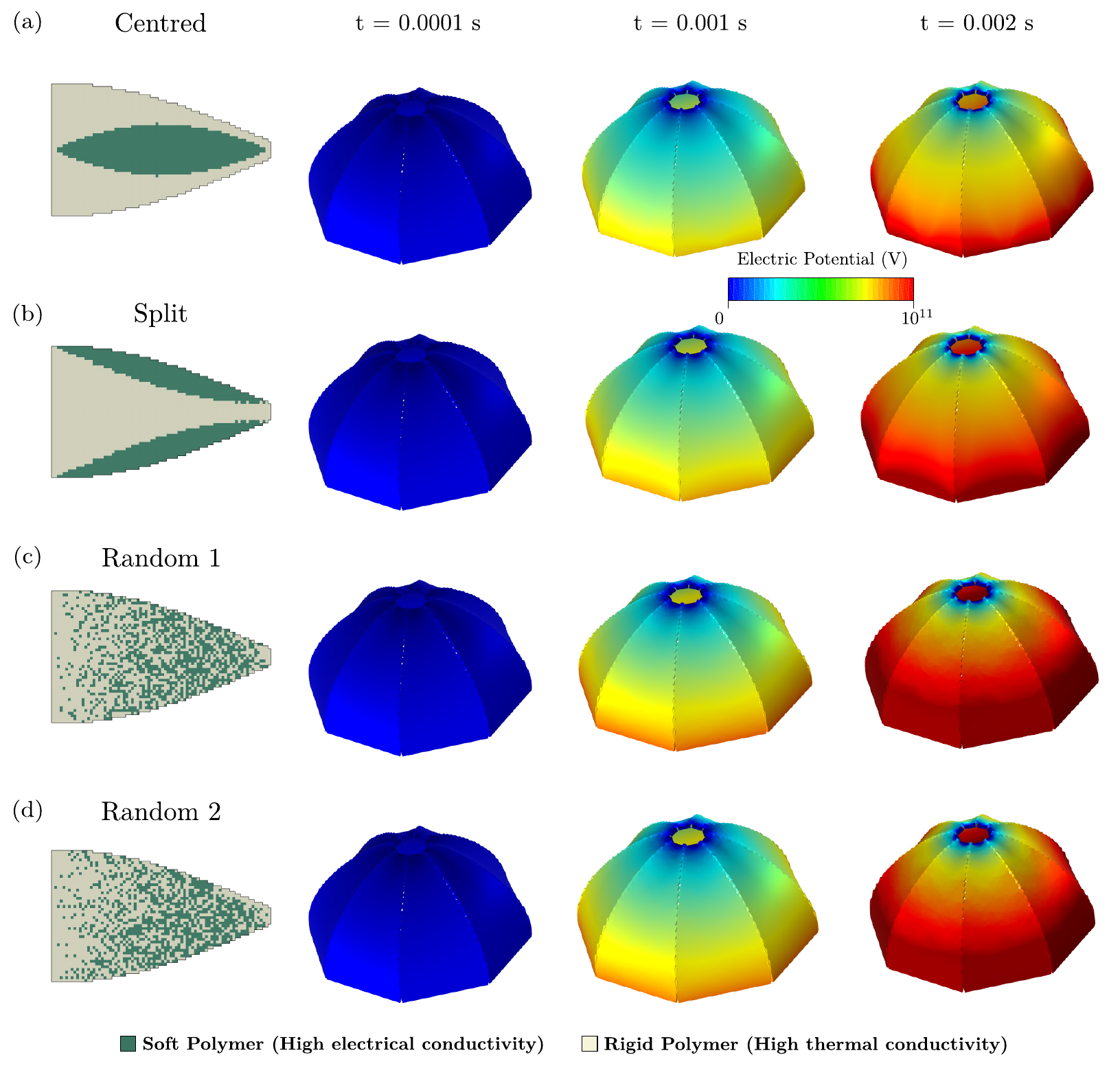}   \caption{\label{fig:hemisphere_electric_potential_images_full_hemisphere.pdf} The contour plots demonstrate the electric currents that arise across deformed hemispherical structures. In the initial column, the material composition of each hemispherical strip is displayed, while the subsequent columns exhibit the electric potential contours at time intervals of t $= 0.0001, 0.001, 0.002$ s.}
\end{figure}

\newpage

\subsubsection{Effective conductivities of FGCs}
\label{section: Effective conductivities of FGCs}

\begin{figure}[!h] 
   \centering
   \includegraphics[width=1\textwidth]{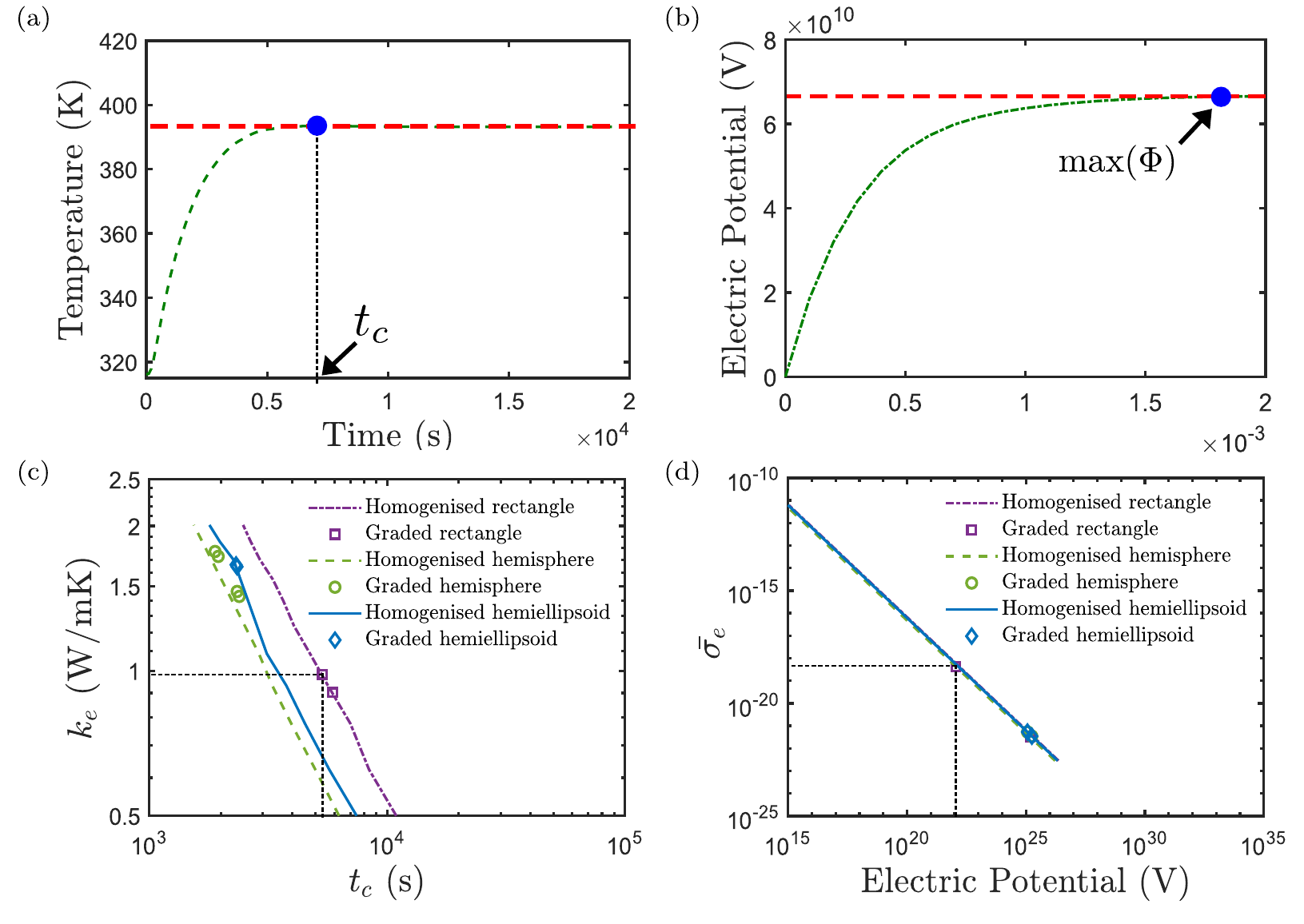}
   \caption{\label{fig: effective_conductivities_explainationd} {Methodology used to determine the effective thermal conductivity, $k_e$, and effective electrical conductivity, $\bar{\sigma_e}$. (a) The definition of critical time, $t_c$, whereby the red dashed line indicates the temperature threshold reaching $99.99\%$ of the equilibrium temperature. The blue point indicates the critical time at which the temperature of the composite exceeds this threshold. (b) The definition of maximum electric potential, max$(\Phi)$, whereby the red dashed line indicates the threshold reaching $99.99\%$ of the maximum electric potential. The relationship between (c) $k_e$ and $t_c$, as well as (d) $\bar{\sigma_e}$ and max$(\Phi)$, for different structures made of single material or FGCs with different geometries.}}
\end{figure}

{To estimate the effective thermal and electrical conductivities of FGCs, we utilise the Fourier's law (Eq. \ref{equ: 3D heat conduction equation}) and the Ohm's law (Eq. \ref{equ:  macroscopic Ohm's law}), respectively. Specifically, since we know the thermal conductivity ($k_e$) determines how fast the temperature field reaches equilibrium, we can define a critical time, $t_c$, according to the numerical result from COMSOL for each FGC structure, to measure this process, which can be correlated it to the corresponding thermal conductivity, $k_e$. As shown in Fig.~\ref{fig: effective_conductivities_explainationd}(a), we define $t_c$ as the time when the temperature at the measure point reaches 99.9\% of the temperature applied at source. For a given structure, according to the Fourier's law, we can notice that $k_e$ is inversely proportional to $t_c$ (i.e. the negative linear relation in the log-log plot, see Fig.~\ref{fig: effective_conductivities_explainationd}(c)). It should be noted that the pre-factors of these linear lines are different for structures with different geometries, which can be obtained by preforming systematic numerical simulations for structures made of uniform materials as benchmarks. Therefore, for each FGC structure with particular geometry, once we measured the value of $t_c$ from simulation results in COMSOL, the effective thermal conductivity $k_e$ can be obtained by mapping to the corresponding benchmark line in Fig.~\ref{fig: effective_conductivities_explainationd}(c). This method enables the determination of the effective thermal conductivities of FGC structures.}

{The similar method can be applied to determine the effective electrical conductivities, $\bar{\sigma_e}$, of FGC structures. However, instead of measuring the critical time of reaching equilibrium, here we correlate the maximum electric potential, max$(\Phi)$, and the electrical conductivity, $\bar{\sigma_e}$, for give structure, according to the Ohm’s law. As illustrated in Fig.~\ref{fig: effective_conductivities_explainationd}(b), max$(\Phi)$ is identified when the plateau of the electric potential is reached, where a threshold of 0.1\% change of the value is set. By preforming numerical simulation in COMSOL for uniform structures with given geometries, the negative linear relation between $\bar{\sigma_e}$ and max$(\Phi)$ in the log-log plot can be determined as benchmarks (see Fig.~\ref{fig: effective_conductivities_explainationd}(d)). Again, preforming simulations for structures made of FGCs with particular geometries, one can easily measure the values of max$(\Phi)$ and determine $\bar{\sigma_e}$ accordingly by mapping to the benchmark lines.}

\end{document}